\documentclass[fleqn,useAMS,usenatbib,usegraphicx]{mn2e}

\usepackage{color}
\usepackage{graphicx}
\usepackage{amsmath}
\usepackage{amssymb}
\usepackage{times}
\usepackage{booktabs}
\usepackage{mathptmx}

\newcommand{\prd}{Physical Review D}
\newcommand{\mnras}{MNRS}

\newcommand{\apj}{Astrophysical Journal}
\newcommand{\apjl}{Astrophys. J.}
\newcommand{\ssr}{Space Science Reviews}
\newcommand{\aap}{Astronomy and Astrophysics}

\title{Modeling magnetized neutron stars using resistive MHD}

\author[Palenzuela]{Carlos~Palenzuela${}^1$ 
\\
${}^1$Canadian Institute for Theoretical Astrophysics, Toronto, Ontario M5S 3H8, Canada
\\}
\pagerange{\pageref{firstpage}--\pageref{lastpage}}

\pubyear{2012}

\begin{document}
%%%%%%%%%%%%%%%%%%%%%%%%%%%%%%%

\maketitle

\label{firstpage}

%%%%%%%%%%%%%%%%%%%%%%%%%%%%%%%

\begin{abstract}
This work presents an implementation of the {\em resistive MHD} equations
for a generic algebraic Ohm's law which includes the effects of finite
resistivity within full General Relativity. The implementation naturally
accounts for magnetic-field-induced anisotropies and, by adopting a
phenomenological current, is able to accurately describe electromagnetic
fields in the star and in its magnetosphere. We illustrate the application
of this approach in interesting systems with astrophysical implications;
the aligned rotator solution and the collapse of a magnetized rotating
neutron star to a black hole.\\
\end{abstract}


\maketitle

%%%%%%%%%%%%%%%%%%%%%%%%%%%%%%%%%%%%%%%%%%%%%%%%%%%%%%%%%%%%%%%%%%%%%%%%%%%%%%%
%%%%%%%%%%%%%%%%%%%%%%%%%%%%%%%%%%%%%%%%%%%%%%%%%%%%%%%%%%%%%%%%%%%%%%%%%%%%%%%
\section{Introduction}
\label{sec:intro}
%%%%%%%%%%%%%%%%%%%%%%%%%%%%%%%%%%%%%%%%%%%%%%%%%%%%%%%%%%%%%%%%%%%%%%%%%%%%%%%

Magnetic fields play an important role in the dynamics of 
many relativistic astrophysical systems such as pulsars, magnetars,
gamma-ray burst (GRBs) and active galactic nuclei (AGNs). 
In many of these scenarios, the Ohmic diffusion timescales of the magnetized
plasma is much longer than the characteristic dynamical timescale of
the system, so one can formally take the limit of infinite electrical
conductivity. This is regarded as the ideal MHD limit, and it is in
general a good approximation to describe astrophysical plasmas. 
Furthermore, such a limit is described by a relatively manageable, but certainly
involved hyperbolic system of equations without stiff terms which facilitates
its computational implementation.
The ideal MHD limit has been extensively used in the last years to study
many of the previous systems (i.e., which basically consist of magnetized
neutron stars and black hole accretion disks) in the fully non-linear regime. 

In spite of its success and convenience, the ideal MHD approximation also has
some limitations. At a purely theoretical level, the assumption of vanishing
electrical resistivity prevents some important physical phenomena such as
dissipation and reconnection of the magnetic field lines. Reconnection
efficiently converts magnetic energy into heat and kinetical energy in
very short timescales. This process is believed to be the mechanism originating
many energetic emissions, such as in soft gamma-ray repeaters (which could be explained by
giant magnetar flares), the Y-point of pulsar magnestosphere or even the
short Gamma-Ray Bursts~\citep{2011SSRv..160...45U}. In order to describe such
processes, schemes going beyond the ideal MHD limit are required.

At the numerical level, all numerical schemes inherit
some numerical resistivity which depends strongly on the resolution, making
difficult to disentangle physical phenomena from numerical artifacts especially
in highly demanding computational scenarios.
The presence of magnetic fields demands relatively
high resolution to accurately capture all the physical processes involved,
many of them occurring at very small scales.
This high resolution is particularly important in the case
of instabilities which amplify the magnetic field, such as
the Kelvin-Helmholtz instability occurring during the merger of binary
neutron stars~\citep{2006Sci...312..719P,2010A&A...515A..30O}, and
the Magneto-Rotational Instability (MRI) occurring in accretion
disks~\citep{1991ApJ...376..214B,1991ApJ...376..223H,
1995ApJ...440..742H,1998RvMP...70....1B}.
Accurate modeling of the rarefied magnetospheres of compact objects similarly
requires high resolution. The electromagnetic fields in this region may
be easier to model by adopting a different limit of the MHD equations
known as the {\em force-free limit}~\citep{Goldreich:1969sb}. 
In this approximation the fluid inertia is neglected, implying that the fluid does 
not influence directly the dynamics of the electromagnetic fields. 

One possibility to overcome these limitations is to consider instead
the resistive MHD framework and solve the full Maxwell and
hydrodynamic equations. The coupling between these two is provided by
the current --by a suitable Ohm's law--. With a convenient choice of
current, including  both induction and Ohmic terms, it is possible
to recover both the ideal MHD limit, as well as the finite-resistivity scheme
required to describe physical dissipation and reconnections. The effect of
small-scales-dynamics can also be modeled with moderate resolutions by using
a suitable current. Finally, magnetically dominated magnetospheres can
be described by a phenomenological current that decouples the fluid from the
force-free EM fields.

The numerical evolution of this resistive MHD code is not free of
difficulties.
The resistive MHD equations can be regarded as an hyperbolic system with
relaxation terms that become stiff for some limits of the current.
Consequently, numerical evolution of this system represents a numerical challenge,
and several works have recently explored different possibilities to
implement it~\citep{Komissarov2007,
2009MNRAS.394.1727P,2009JCoPh.228.6991D,2010ApJ...716L.214Z,
2011ApJ...735..113T,2012arXiv1205.2951B,2012arXiv1208.3487D}.
In this work we take a step further in the development of one of
these approaches, based on the Implicit-Explicit (IMEX) Runge-Kutta.
Our aim is to model both the interior and the exterior of a star with
a phenomenological current based on physical arguments. This will be
particularly interesting to study the electromagnetic emissions of
astrophysical relativistic systems involving magnetized neutron stars.

The capabilities of our approach are tested by considering 
the force-free aligned rotator solution,
a well studied problem in the context of pulsar magnetospheres
~\citep{2006ApJ...643.1139C,2006ApJ...648L..51S,McKinney:2006sd,2006MNRAS.368.1717B,
2009A&A...496..495K,2012ApJ...746...60L,2012arXiv1211.2803T}. These works were 
restricted to flat spacetime and excluded the interior of the star from the computational
domain, thus side-stepped
the stiffness problem mentioned above. Recently a hybrid approach, matching both the ideal
and force-free system of equations, revisited this problem within a framework capable
of studying both star and surrounding magnetosphere within General
Relativity~\citep{2011arXiv1112.2622L}. 
However such a scheme still relies on two different approximations applied in two regions.
The approach we present here finally allows for treating the system from a global point of
view with a single, general relativistic, framework. 

We also revisit another important astrophysical scenario with a
much less understood dynamics; the collapse of a magnetized neutron star
to a black hole. This system represents even a more challenging problem
because of the strong gravity fields, and it has been studied numerically
by considering different  approximations. An early study matched an analytical
solution for the star to an electrovacuum magnetosphere~\citep{Baumgarte:2002b}.
More recently, further realism was achieved by adopting the hybrid scheme that matched 
the numerical solution of the star to force-free  magnetosphere~\citep{2011arXiv1112.2622L}.
A step towards studying this system within a common, resistive, framework was presented
in~\citep{2012arXiv1208.3487D}, although the star's exterior was treated as an electrovacuum
magnetosphere. Our approach presented here is able to consistently study the star and
its force-free magnetosphere within the general relativistic resistive MHD equations.

The paper is organized as follows. Section~\ref{sec:equations} 
summarizes the fully relativistic resistive MHD system, which is 
slightly different from the one adopted in~\citep{2012arXiv1208.3487D}.
In Section~\ref{sec:coupling_EM_fluid} it is discussed
a generic family of algebraic Ohm's law, and how to construct
a phenomenological current to recover both the ideal MHD and the
force-free limits. Section~\ref{sec:hyperbolic_relaxation}
summarizes briefly the IMEX Runge-Kutta methods and different
techniques to solve generically the implicit step for any algebraic
form of the relaxation terms. The application of these methods to
the resistive MHD system is performed is section~\ref{sec:rmhd}.
Section~\ref{sec:simulations} presents our numerical results for the
aligned rotator and the collapse of a neutron star to a black hole.
We conclude with some remarks.

Throughout this work we adopt geometric units such that $G=c=1$,
and the convention where greek indices $\mu,\nu,\alpha,...$
denote spacetime components (ie, from $0$ to $3$), while roman indices
$i,j,k,...$ denote spatial ones. Bold letters will represent vectors.

%%%%%%%%%%%%%%%%%%%%%%%%%%%%%%%%%%%%%%%%%%%%%%%%%%%%%%%%%%%%%%%%%%%%%%%%%%%%%%%
\section{The evolution equations}
\label{sec:equations}
%%%%%%%%%%%%%%%%%%%%%%%%%%%%%%%%%%%%%%%%%%%%%%%%%%%%%%%%%%%%%%%%%%%%%%%%%%%%%%%

This section summarizes the general relativistic resistive
magnetohydrodynamic equations, that will allow us to model
self-gravitating magnetized fluids.
The evolution of the spacetime geometry
is governed by Einstein equations. The electromagnetic fields and the
fluid obey, respectively, the Maxwell and the General Relativistic Hydrodynamic equations.
The closure of the system is given by two constitutive equations;
the first one is the equation of state, which relates the pressure
to the other fluid variables. The second one is Ohm's law,
--defining the coupling between the fluid and the electromagnetic fields--
which will be described in the next section.

%%%%%%%%%%%%%%%%%%%%%%%%%%%%%%%%%%%%%%%%%%%%
\subsection{Einstein Equations}
\label{subsec:BSSN}
%%%%%%%%%%%%%%%%%%%%%%%%%%%%%%%%%%%%%%%%%%%%%

The geometry of the spacetime can
be obtained by solving the four-dimensional Einstein equations.
These equations can be recast as a standard initial value problem
by splitting explicitly the time and the space coordinates through
a 3+1 decomposition, such that the line element can be expressed as
\begin{eqnarray}
ds^2 &=& g_{\mu\nu}\, dx^{\mu} dx^{\nu} \nonumber \\
         &=&-\alpha^2 \, dt^2 
          + \gamma_{ij}\left(dx^i + \beta^i\, dt\right)
                       \left(dx^j + \beta^j\, dt\right),
\end{eqnarray}
where $g_{\mu\nu}$ is the spacetime metric, $\gamma_{ij} = g_{ij}$
is the intrinsic metric of the spacelike hypersurfaces, and 
the lapse function $\alpha$ and the shift vector $\beta^i$
relates how the coordinates change between neighboring hypersurfaces.
The normal to the hypersurfaces is given explicitly by
\begin{equation}
 n^{\mu} = \frac{1}{\alpha}(1 ,-\beta^i) ~~~,~~~ 
 n_{\mu} = (-\alpha, 0)  ~~.
\end{equation}
Indices on spacetime quantities are raised and lowered with the 4-metric
and its inverse, while the 3-metric and its inverse
are used to raise and lower indices on spatial quantities.

The rate of change of the intrinsic curvature from
one hypersurface to another is given by the extrinsic curvature
\begin{equation}
  K_{ij} = -\frac{1}{2 \alpha} (\partial_t - {\cal L}_{\beta}) \gamma_{ij}
\end{equation}
where ${\cal L}_{\beta}$ is the Lie derivative along the vector $\beta^i$.

At any given time, the spacetime geometry is then fully defined
by the $3+1$ variables $\{ \alpha,\beta^i,\gamma_{ij},K_{ij} \}$.
We adopt the Baumgarte-Shapiro-Shibata-Nakamura
(BSSN) formulation of Einstein's equations to evolve a suitable
combination of these fields, in a form very close to the 
presented in~\citep{2006PhRvL..96k1101C}.

%%%%%%%%%%%%%%%%%%%%%%%%%%%%%%%%%%%%%%%%%%%%
\subsection{Maxwell equations}
\label{subsec:maxwell}
%%%%%%%%%%%%%%%%%%%%%%%%%%%%%%%%%%%%%%%%%%%%%

The electromagnetic fields follow Maxwell equations,
that in their extended version can be written as~\citep{2010PhRvD..81h4007P} 
\begin{eqnarray}
  \nabla_{\mu} (F^{\mu \nu} + g^{\mu \nu} \psi) &=& - I^{\nu} + \kappa n^{\nu} \psi
\label{Maxwell_ext_eqs2a} \\
 \nabla_{\mu} ({}^*F^{\mu \nu} + g^{\mu \nu} \phi) &=& \kappa n^{\nu} \phi~,
\label{Maxwell_ext_eqs2b} 
\end{eqnarray}
where $\{F^{\mu \nu}, ^*F^{\mu \nu}\}$ are the Maxwell and the Faraday tensors,
$I^{\nu}$ is the electric current and $\{ \phi,\psi \}$ are scalars introduced
to control dynamically the constraints by exponentially damping them
in a characteristic time $1/\kappa$~\citep{2002JCoPh.175..645D}. When both the electric and magnetic susceptibility of the medium
vanish, like in vacuum or in a highly ionized plasma, the Faraday
tensor is simply the dual of the Maxwell one,
\begin{equation}\label{Fduals}
  {}^*F^{\mu \nu} = \frac{1}{2}\, \epsilon^{\mu \nu \alpha \beta} \,
  F_{\alpha\beta}~,\qquad
  F^{\mu \nu} = - \frac{1}{2}\, \epsilon^{\mu \nu \alpha \beta}~{}^*F_{\alpha \beta}
\end{equation}
where $\epsilon^{\mu \nu \alpha \beta}$ is the Levi-Civita
pseudotensor of the spacetime, related to the 4-indices
Levi-Civita symbol $\eta^{\mu \nu \alpha \beta}$ by
\begin{equation}\label{levicivita}
   \epsilon^{\mu \nu \alpha \beta} = \frac{1}{\sqrt{g}}~
   \eta^{\mu \nu \alpha \beta} \qquad
    \epsilon_{\mu \nu \alpha \beta} = -\sqrt{g}~
   \eta_{\mu \nu \alpha \beta}~.
\end{equation}
In this case, both tensors can be decomposed in terms of the
electric and magnetic fields,
\begin{eqnarray}
  F^{\mu \nu} &=& n^{\mu} E^{\nu} - n^{\nu} E^{\mu} 
               + \epsilon^{\mu\nu\alpha\beta}~B_{\alpha}~n_{\beta} 
\label{F_em1a} \\
  {}^*F^{\mu \nu} &=& n^{\mu} B^{\nu} - n^{\nu} B^{\mu} 
               - \epsilon^{\mu\nu\alpha\beta}~E_{\alpha}~n_{\beta}
\label{F_em1b}
\end{eqnarray}
such that $E^{\mu}$ and $B^{\mu}$ are the electric and magnetic
fields measured by a normal observer $n^{\mu}$. Both fields
are purely spatial, that is, $E^{\mu} n_{\mu} = B^{\mu} n_{\mu} = 0$. 

The covariant Maxwell equations (\ref{Maxwell_ext_eqs2a},
\ref{Maxwell_ext_eqs2b}) can be written, by performing the 3+1
decomposition, in term of the electromagnetic fields and the
divergence-cleaning scalars~\citep{2010PhRvD..81h4007P} as,
\begin{eqnarray}
  (\partial_t - {\cal L}_{\beta}) E^{i} &-& \epsilon^{ijk} \nabla_j (\alpha B_k) 
   + \alpha \gamma^{ij} \nabla_j \psi 
\\ &=& \alpha trK E^i - \alpha J^i 
\nonumber 
\label{maxwellext_3+1_eq1a} \\
  (\partial_t - {\cal L}_{\beta}) \psi &+& \alpha \nabla_i E^i = 
    \alpha q -\alpha \kappa \psi
\label{maxwellext_3+1_eq1b} \\
  (\partial_t - {\cal L}_{\beta}) B^{i} &+& \epsilon^{ijk} \nabla_j (\alpha E_k) 
   + \alpha \gamma^{ij} \nabla_j \phi
\\ &=& \alpha trK B^i 
\nonumber
\label{maxwellext_3+1_eq1c} \\
  (\partial_t - {\cal L}_{\beta}) \phi &+& \alpha \nabla_i B^i = 
   -\alpha \kappa \phi ~~.
\label{maxwellext_3+1_eq1d}
\end{eqnarray}
where $\epsilon^{ijk} \equiv \epsilon^{ijk\alpha} n_{\alpha} =
\eta^{ijk}/\sqrt{\gamma}$ is the three-dimensional Levi-Civita pseudotensor.
Since $F^{\mu \nu}$ is antisymmetric, the four-divergence of
equation (\ref{Maxwell_ext_eqs2a}) leads to an additional 
equation for the current conservation of Maxwell solutions,
\begin{equation}\label{conserved_current}
   \nabla_{\mu} I^{\,\mu} = 0 ~.
\end{equation}
The electric current $I^{\nu}$ can be decomposed into components
along and perpendicular to the vector $n^{\nu}$,
\begin{equation}\label{current_decomposition}
  I^{\nu} = n^{\nu} q + J^{\nu}~~,
\end{equation}
where $q$ and $J^{\nu}$ are the charge density and the current as observed
by a normal observer $n^{\nu}$. Again, $J^{\nu}$ is purely spatial,
so $J^{\nu} n_{\nu}=0$. The current conservation (\ref{conserved_current})
can be expressed, with the 3+1 decomposition, as
\begin{equation}\label{consJ_3+1} 
  (\partial_t - {\cal L}_{\beta}) q + \nabla_i (\alpha J^i) = \alpha trK q
\end{equation}
Only a prescription for the spatial components $J^i$, which will determine
the coupling between the EM fields and the fluid, is required to
complete the system of Maxwell equations. This relation, 
commonly known as Ohm's law, will be discussed in detail in
section~\ref{sec:coupling_EM_fluid}.

%%%%%%%%%%%%%%%%%%%%%%%%%%%%%%%%%%%%%%%%%%%%%%%
\subsection{Hydrodynamic equations}
\label{subsec:hydro}
%%%%%%%%%%%%%%%%%%%%%%%%%%%%%%%%%%%%%%%%%%%%%%%

A perfect fluid minimally coupled to an electromagnetic field
is described by the total stress-energy tensor
\begin{eqnarray}\label{stress-energy-perfectfluid}
   T_{\mu \nu} &=& \left[ \rho (1 + \epsilon) + p \right]
                    u_{\mu} u_{\nu} + p g_{\mu \nu}
\nonumber \\
   &+& {F_{\mu}}^{\lambda} F_{\nu \lambda}
     - \frac{1}{4} g_{\mu \nu} ~ F^{\lambda \alpha} F_{\lambda \alpha} 
\end{eqnarray}
where a factor $1/\sqrt{4\pi}$ has been absorbed in the definition 
of the electromagnetic fields. Here $\rho$ is the rest mass density,
$\epsilon$ the internal energy and $p$ is the pressure, given by a
closure relation $p=p(\rho,\epsilon)$ commonly known as the equation
of state (EoS). These fluid quantities are measured in the rest frame
of the fluid element. However, to describe the system is usually more
convenient to adopt an Eulerian perspective where coordinates are
not tied to the flow of the fluid. The four-velocity $u^{\mu}$ describes
how the fluid moves with respect to the Eulerian observers, and 
can be decomposed into space and time components,
\begin{equation}\label{velocity_decomposition}
   u^{\mu} = W \left( n^{\mu} +  v^{\mu} \right)
\end{equation}
where $v^{\mu}$ corresponds to the familiar three-dimensional
velocities as measured by Eulerian observers (i.e., $v^{\mu} n_{\mu} = 0$).
The time component is defined by the
normalization relation $u^{\mu} u_{\mu} = -1$, such that
\begin{equation}\label{velocity_decomposition2}
   W = - n_{\mu} u^{\mu} = (1 - v_i v^i)^{-1/2} ~,~~
%  u^{i} = W (v^{i} - \frac{\beta^i}{\alpha})
\end{equation}
where we can now recognize $W$ as the Lorentz factor.

In summary, the magnetized fluid is described by the
physical fields (i.e., the fluid variables and the electromagnetic
fields) plus the divergence cleaning scalars, which form
the set of primitive variables 
$(\rho, \epsilon, p, v^i, E^i, B^i, q, \phi, \psi)$
The matter evolution must comply with the conservation of
the total stress-energy tensor
\begin{equation} \label{Tconserv_eq}
  \nabla_{\nu}T^{\mu\nu}= 0,
\end{equation}
which can be expressed as a system of conservation laws for
the energy density $U$ and the momentum density $S_i$, 
defined from the projections of the stress-energy tensor
\begin{equation}\label{Tmunu_projections}
   U = n_{\mu} n_{\nu} T^{\mu \nu}  ~~,~~
   S_{i} =  - n^{\mu} T_{\mu i} ~~,~~ 
   S_{ij} = T_{ij} ~~.
\end{equation}
In addition to the conservation of energy and momentum,
the fluid usually also conserves the total number of particles,
\begin{equation}\label{baryon_conservation}
  \nabla_{\mu} ( \rho u^{\mu} ) = 0 
\end{equation}
where $\rho u^{\mu}$ is the baryon number density. This equation
is just the relativistic generalization of the conservation of mass.

As mentioned above, it is necessary to specify the EOS
to define the  pressure and complete the system of hydrodynamic
equations. Along this paper we will consider either the polytropic
EoS $p = K \rho^{\Gamma}$, which is a good approximation to describe
cold stars, and the ideal gas EoS $p = (\Gamma -1) \rho \epsilon$,
which allows for shock heating in the fluid.

%%%%%%%%%%%%%%%%%%%%%%%%%%%%%%%%%%%%%%%%%%%%%%%
\subsection{Resistive MHD system}
\label{subsec:finaleqs}
%%%%%%%%%%%%%%%%%%%%%%%%%%%%%%%%%%%%%%%%%%%%%%%

The evolution of the electromagnetic fields follows the Maxwell
equations and the conservation of charge, while the fluid
fields are governed by the conservation of the total energy,
momentum and baryonic number. In order to capture accurately
the weak solutions of these non-linear equations in presence of
shocks it is important to express them as a set of local conservation
laws, namely
\begin{eqnarray}
\label{maxwell11}
  \partial_t  (\sqrt{\gamma} B^i) 
  &+& \partial_k [\sqrt{\gamma} \left( -\beta^k B^i 
   + \alpha ( \epsilon^{ikj} E_j + \gamma^{ik} \phi ) \right) ] 
\\
   &=&
   - \sqrt{\gamma} B^k (\partial_k \beta^i) 
   + \sqrt{\gamma} \phi \left( \gamma^{ij} \partial_j \alpha
   - \alpha \gamma^{jk} \Gamma^i_{jk} \right)
\nonumber \\
\label{maxwell12}
  \partial_t  (\sqrt{\gamma} E^i) 
  &+& \partial_k [\sqrt{\gamma} \left( -\beta^k E^i 
   - \alpha ( \epsilon^{ikj} B_j - \gamma^{ik} \psi ) \right) ] 
\\
   &=&
   - \sqrt{\gamma} E^k (\partial_k \beta^i) 
   + \sqrt{\gamma} \psi \left( \gamma^{ij} \partial_j \alpha
   - \alpha \gamma^{jk} \Gamma^i_{jk} \right)
\nonumber \\
  && - \alpha \sqrt{\gamma} J^i 
\nonumber \\
\label{maxwell13}
  \partial_t  (\sqrt{\gamma} \phi) &+& \partial_k [\sqrt{\gamma}
   (- \beta^k \phi + \alpha B^k)] 
\\
   &=& 
   \sqrt{\gamma}[ - \alpha\,  \phi\, trK + B^k (\partial_k \alpha) 
   -\alpha \kappa \phi] 
\nonumber \\
\label{maxwell14}
  \partial_t (\sqrt{\gamma} \psi) &+& \partial_k 
      [\sqrt{\gamma} (- \beta^k \psi + \alpha E^k)] 
\\
    &=& 
   \sqrt{\gamma}[- \alpha\,  \psi\, trK + E^k (\partial_k \alpha) 
   + \alpha q -\alpha \kappa \psi] 
\nonumber \\
\label{maxwell15}
  \partial_t  (\sqrt{\gamma} q) 
  &+& \partial_k [\sqrt{\gamma} (- \beta^k q + \alpha J^k)] = 0 \\
\label{fluid11}
  \partial_t  (\sqrt{\gamma} D ) 
   &+& \partial_k [\sqrt{\gamma} (- \beta^k + \alpha v^k) D ] = 0 \\
\label{fluid12}
  \partial_t (\sqrt{\gamma} \tau) 
   &+& \partial_k [\sqrt{\gamma} \left( - \beta^k \tau + \alpha (S^k - v^k D)  \right)] 
\\
   &=& \sqrt{\gamma} [\alpha S^{ij} K_{ij} - S^j \partial_j \alpha] \\
\label{fluid13}
  \partial_t (\sqrt{\gamma} S_i) 
   &+& \partial_k [\sqrt{\gamma} (- \beta^k S_i + \alpha {S^k}_i)] 
\\
   &=& \sqrt{\gamma} [\frac{\alpha}{2} S^{jk} {\partial_{i}} \gamma_{jk} 
                   + S_j \partial_i \beta^j - (\tau + D) \partial_i \alpha] 
\nonumber
\end{eqnarray}
where we have defined
\begin{eqnarray}\label{finalsets}
    D &=& \rho W  ~,~~~ \\
   \tau &=& h W^2 - p + \frac{1}{2} (E^2 + B^2) - \rho W ~,~~~ \\
   S_{i} &=& h W^2 v_{i} + \epsilon_{ijk} E^j B^k ~,~~~ \\
   S_{ij} &=& h W^2 v_{i} v_{j} + \gamma_{ij} p \\
          && - E_i E_j - B_i B_j 
             + \frac{1}{2} \gamma_{ij} (E^2 + B^2) ~,~~
\nonumber
\end{eqnarray}
and the enthalpy $h \equiv \rho (1 + \epsilon) + p$.
This form of the relativistic resistive MHD equations is
basically the same presented already in~\citep{2012arXiv1208.3487D}.
Another similar formulation has also been derived
recently~\citep{2012arXiv1205.2951B}.
Notice also that the energy conservation has been expressed in terms
of the quantity $\tau \equiv U - D$ to recover the
Newtonian limit of the energy density.

%%%%%%%%%%%%%%%%%%%%%%%%%%%%%%%%%%%%%%%%%%%%%%%%%%%%%%%%%%%%%%%%%%%%%%
\section{Coupling between the EM fields and the fluid}
\label{sec:coupling_EM_fluid}
%%%%%%%%%%%%%%%%%%%%%%%%%%%%%%%%%%%%%%%%%%%%%%%%%%%%%%%%%%%%%%%%%%%%%%

Maxwell and hydrodynamic equations are coupled by
means of the current ${I}^{\mu}$, whose explicit form 
generically depends on the electromagnetic fields and the local
fluid properties measured in the comoving frame. Consequently,
it is convenient to introduce the electric and magnetic fields 
measured by an observer comoving with the fluid, namely
$e^{\mu} \equiv F^{\mu \nu} u_{\nu}$ and $
b^{\mu} \equiv {}^*F^{\mu \nu} u_{\nu}$.
Notice that, since $e^{\mu} u_{\mu} = b^{\mu} u_{\mu} = 0$,
there are only three independent components. The Maxwell and Faraday
tensors can therefore be expressed as
\begin{eqnarray}
  F^{\mu \nu} &=& u^{\mu} e^{\nu} - u^{\nu} e^{\mu} 
               + \epsilon^{\mu\nu\alpha\beta}~b_{\alpha}~u_{\beta} 
\label{F_em2a} \\
  {}^*F^{\mu \nu} &=& u^{\mu} b^{\nu} - u^{\nu} b^{\mu} 
               - \epsilon^{\mu\nu\alpha\beta}~e_{\alpha}~u_{\beta}
\label{F_em2b}
\end{eqnarray}
and the electric current can be decomposed into components along 
and transverse to $u^{\nu}$,
\begin{equation}\label{current_decomposition2}
  I^{\mu} = u^{\mu} \tilde{q} + j^{\mu}~~,
\end{equation}
where $j^{\mu} u_{\mu} = 0$ and $\tilde{q}$ is the charge density
measured by the comoving observer. The relation with the Eulerian
quantities (\ref{current_decomposition}) can be obtained from
\begin{equation}
 q = -n_{\mu} I^{\mu} = W \tilde{q} - n_{\mu} j^{\mu}~~. 
\end{equation}
Substituting these results into eq.~(\ref{current_decomposition2}) and
using the 3+1 decomposition of the four-velocity, one can write
the spatial components of the current as
\begin{equation}\label{current_decomposition3}
  I_{i} = J_i = (q + j^{\mu} n_{\mu}) v_i + j_i ~~.
\end{equation}
Since the charge density follows directly from the current
conservation~(\ref{consJ_3+1}), the prescription for the
three-dimensional electrical current $J_i$ is the only
missing piece to completely determine Maxwell equations.

%%%%%%%%%%%%%%%%%%%%%%%%%%%%%%%%%%%%%%
\subsection{Generalized covariant Ohm's law}
\label{sec:covohmlaw}
%%%%%%%%%%%%%%%%%%%%%%%%%%%%%%%%%%%%%%

A standard prescription, known as the Ohm's law,
is to consider that the current is proportional to
the Lorentz force acting on a charged particle,
implying a linear relation between the current and the
electric field in the comoving frame.
A richer variety of physical phenomena may be described
by including also additional terms proportional to the comoving
magnetic field, leading to a generalized covariant
Ohm's law of the form,
\begin{equation}\label{ohm_law_covariant}
 j^{\mu} = \sigma^{\mu \nu} e_{\nu} + \lambda\, b^{\mu}~,
\end{equation}
being $\sigma^{\mu \nu}$ the electrical conductivity of
the medium~\citep{1978PhRvD..18.1809B} and 
$\lambda$ a parameter related to the covariant generalization
of the mean-field dynamo~\citep{2012arXiv1205.2951B}.

The electrical conductivity can be calculated either 
in the collision-time approximation~\citep{1978PhRvD..18.1809B} 
or in the framework of relativistic charged multifluids
~\citep{2012arXiv1204.2695A}, leading to the same main results.
The tensorial conductivity can be written as,
\begin{equation}\label{tensorial_conductivity}
 \sigma^{\mu \nu} =  \frac{\sigma}{1 + \xi^2 b^2} ( g^{\mu \nu} 
    + \xi^2 b^{\mu} b^{\nu}
    + \xi \epsilon^{\mu \nu \alpha \beta} u_{\alpha} b_{\beta} )  ~~
\end{equation}
where the coefficients are given by
\begin{equation}\label{tensorial_conductivity2}
  \xi = 1/R = e \tau_r / m_e ~~,~~
  \sigma = R / (n_e e) ~~.
\end{equation}
Here $\tau_r$ is the collision or relaxation time, $n_e$ is the
electron density and $e$ and $m_e$ are the electron's charge and mass.
In the framework described in~\citep{2012arXiv1204.2695A}, 
$R$ is introduced as a proportionality constant in the dissipative
force between the two components of the fluid.
It is easy to check that the first term of the conductivity
(\ref{tensorial_conductivity}) leads to the well known isotropic
scalar case, while the other two represent the anisotropies
due to the presence of a magnetic field, corresponding to the 
Hall effect.

In order to compute the closure relation (\ref{current_decomposition3})
it is necessary to write the general relativistic Ohm's law
in terms of fields measured by an Eulerian observer. 
Let us first consider a simplified Ohm's law neglecting both
the dynamo effects and the last term in the tensorial 
conductivity (\ref{tensorial_conductivity}),
\begin{equation}\label{complete_tensorial_conductivity}
 j_{\mu} =  \frac{\sigma}{1 + \xi^2 b^2} [ e_{\mu} 
         +  \xi^2 (e_{\nu} b^{\nu}) b_{\mu} ]~~,
\end{equation}
as it has also been used in~\citep{2011MNRAS.418.1004Z}.
It was pointed out that this current implies an incomplete Hall
effect~\citep{2012arXiv1204.2695A}, but it will be enough
for our later discussion. Within these assumptions, and using
that the electric and magnetic fields in the fluid frame can be
written as
\begin{eqnarray}\label{em_fluid_frame}
  e^{\mu} &=&  W n^{\mu} (E^{\nu} v_{\nu}) + W E^{\nu}
           + W \epsilon^{\mu \nu \alpha} v_{\nu} B_{\alpha}
\label{relation_eE} \\
  b^{\mu} &=&  W n^{\mu} (B^{\nu} v_{\nu}) + W B^{\nu}
           - W \epsilon^{\mu \nu \alpha} v_{\nu} E_{\alpha} ~~,
\label{relation_bB}
\end{eqnarray}
it is straightforward to obtain the contraction
\begin{eqnarray}\label{contracted_jn}
 j_{\mu} n^{\mu} &=&  \frac{\sigma}{1 + \xi^2 b^2} [ e_{\mu} n^{\mu} 
         +  \xi^2 (e_{\nu} b^{\nu}) b_{\mu} n^{\mu} ]
\\
     &=& \frac{\sigma}{1 + \xi^2 b^2} [-W (E^k v_k) 
                      - W \xi^2 (E^j B_k) (B^k v_k) ] ~~.
\nonumber
\end{eqnarray}
The prescription for the spatial current (\ref{current_decomposition3})
can be now computed, leading to
\begin{equation}\label{ohm_relativistic_spatial}
 J_i =  q v_i + \frac{\sigma}{1 + \xi^2 b^2} 
         [{\cal E}_i + \xi^2 (E^k B_k) {\cal B}_i]
\end{equation}
where we have introduced the shortcuts
\begin{eqnarray}\label{newE}
 {\cal E}_i &=& W  \left[ E_i + \epsilon_{i j k} v^{j} B^{k}
                         - (v_{k} E^{k}) v_{i} \right] ~,~~
\\
\label{newB}
 {\cal B}_i &=& W \left[ B_i - \epsilon_{i j k} v^{j} E^{k}
                         - (v_{k} B^{k}) v_{i} \right] ~.
\end{eqnarray}
It is important to recall that this current accounts not only
for isotropic resistivity but also for some anisotropic
effects induced by the magnetic fields.

In the regime of low magnetization (i.e., $p/B^2 \gg 1$) 
these anisotropic effects are expected to be small,
implying $\xi \ll 1$. In this limit the
third term in the current (\ref{ohm_relativistic_spatial}) can
be neglected, leading
to the well-known isotropic Ohm's law. The high conductivity of
the fluid implies that, in order to get a finite current,
the electric field measured by the comoving observers must
vanish
\begin{equation}\label{ef_imhd}
   e^{\mu} = 0 \longrightarrow E^i = - \epsilon^{ijk} v_j B_k \,.
\end{equation}
This is the ideal-MHD condition, which states that the electric
field is not an independent variable since it can be obtained
via a simple algebraic relation from the velocity and the
magnetic vector fields.

The anisotropic effects are expected to be important in 
magnetically dominated fluids (i.e., $p/B^2 \ll 1$). In this limit
$\xi \gg 1$, and the second term in the current
(\ref{ohm_relativistic_spatial}) can be neglected. In highly
conducting fluids a finite current is recovered only if
the electric field is perpendicular to the magnetic field,
\begin{equation}\label{ef_iff}
   e^{\mu} b_{\mu} = E^i B_i = 0 ~.
\end{equation}
since the initial assumption of magnetically dominated fluid
prevents the trivial solution $b^i=0$. In the next subsection it
will be shown that this relation is one of the constraints of the
force-free approximation.

%%%%%%%%%%%%%%%%%%%%%%%%%%%%%%%%%%%%%%
\subsection{The force-free limit}
\label{subsec:currentforcefree}
%%%%%%%%%%%%%%%%%%%%%%%%%%%%%%%%%%%%%%

The magnetospheres of magnetized neutron stars~\citep{Goldreich:1969sb}
and black holes immersed in externally sourced magnetic
fields~\citep{Blandford1977} are filled with a low-density plasma so
rarefied that even moderate magnetic fields stresses can easily dominate over
the pressure gradients.  In this regime, the main contribution to 
the stress-energy tensor comes from the electromagnetic
part, $T_{\mu \nu} \approx T_{\mu \nu}^{em}$. Allowing by Maxwell
equations, the total conservation of energy and momentum can be written
as
\begin{equation}\label{div-stress-fluidem2}
   0 = \nabla_{\nu} T^{\mu \nu} \approx - F^{\mu \nu} I_{\nu} ~~.
\end{equation}
The vanishing of the Lorentz force $F^{\mu \nu} I_{\nu}$ leads
to an approximation known as force-free limit, which is valid
only for magnetically dominated plasmas with negligible inertia.
The spatial components of the force-free condition 
~(\ref{div-stress-fluidem2}), after performing the 3+1 decomposition, are
\begin{equation}\label{forcefree2}
   q E^i + \epsilon^{ijk} J_j B_k = 0 ~~
\end{equation}
or, after some simple manipulations,
\begin{equation}\label{forcefree3}
   J^i = q\, v^i_{d} + (J^k B_k) \frac{B^i}{B^2} ~~~,~~~
   E^i B_i = 0 ~~,
\end{equation}
where we have defined
$v^i_{d} \equiv \epsilon^{ijk} E_j B_k/{B^2}$
as the drift velocity.
Several options have been proposed to compute the
term $J^k B_k$, which is crucial to provide a completely explicit relation
for the current.
For instance, a closed formed for the current can be calculated by
enforcing the constraint $\partial_t (E^i B_i) = 0$~\citep{2007ApJ...667L..69G}.
Another option is to evolve Maxwell equations by considering
only the drift term of the current (\ref{forcefree3}), and 
correct the electric field after each timestep to satisfy the 
other force-free condition $E^i B_i = 0$~\citep{2004MNRAS.350..427K,2006ApJ...648L..51S}.
This approximation has been used successfully to study numerically
pulsar magnetospheres \citep{2006ApJ...648L..51S} and
jets emerging from black holes with an externally sourced 
magnetic field \citep{2010Sci...329..927P,2010PhRvD..82d4045P,2011PNAS..10812641N}. 

The force-free limit can also be achieved by considering an
effective anisotropic conductivity with a generic 
form given by~\citep{2004MNRAS.350..427K,2012ApJ...749L..32M,2012ApJ...754...36A}
\begin{equation}\label{forcefree4}
      J^i = q\, v^i_{d} + \frac{\sigma_{\parallel}}{B^2}
     \left[  (E^k B_k) B^i + \chi (E^2-B^2) E^i \right] ~~,~~~
\end{equation}
where $\sigma_{\parallel}$ is the (anisotropic) conductivity along
the magnetic field lines. The additional term proportional to $E^2-B^2$
is introduced in order to enforce the physical constraint
$|E|>|B|$. The remarkably close resemblance between the covariant current
(\ref{ohm_relativistic_spatial}) and the force-free one
(\ref{forcefree4}) suggests that both of them could lead to the same
solutions for some limit of the conductivities.
However, the force-free current (\ref{forcefree4}) attains
a particularly interesting feature; due mainly to the assumption of
negligible fluid inertia, it does not depend on the fluid
fields. This means that the EM fields are decoupled to the fluid
variables, an advantage that could be used to model accurately
the EM fields in regions where the fluid description is not accurate.

%%%%%%%%%%%%%%%%%%%%%%%%%%%%%%%%%%%%%%
\subsection{A current for the ideal MHD and the force-free limits}
\label{subsec:current_ideal_ff}
%%%%%%%%%%%%%%%%%%%%%%%%%%%%%%%%%%%%%%

The numerical evolution of the ideal MHD equations
typically fails in low density regions with high magnetization
unless sufficient resolution is available,
a situation that arises commonly in the magnetospheres.
A standard practice to avoid these failures is to maintain
a density floor (i.e., the so called {\em atmosphere}) in regions of low density
to exploit advanced numerical techniques for relativistic hydrodynamics.
The density in the atmosphere is much smaller than that inside the star,
so this approach does not affect the star's dynamics. However, in the 
magnetosphere the fluid inertia (and pressure) is typically much smaller
than that of the electromagnetic field and one generally
encounters numerical difficulties. 
These problems are mitigated by increasing the density in the atmosphere,
effectively decreasing the magnetization in the exterior of the star.
Although these modifications produce an unphysical modeling of the plasma
in the magnetosphere, one could still solve correctly Maxwell equations
by using a suitable current that decouples the electromagnetic fields
from the fluid variables.

As explained earlier, the covariant current 
(\ref{ohm_relativistic_spatial}) reduces to the ideal MHD limit
for high isotropic conductivities (i.e., $\sigma \rightarrow \infty$
and $\xi \rightarrow 0$), while that the force-free constraint
$E^i B_i = 0$ is enforced for large anisotropic conductivities
(i.e., $\sigma,\xi \rightarrow \infty$). This suggests that the solutions
for the EM fields in both limits can be achieved just by changing the
anisotropic conductivity, independently on the plasma magnetization.
Although Ohm's law~(\ref{ohm_relativistic_spatial}) is quite general,
it still couples the EM fields to the velocity.  In addition, the parameter $\xi$ is not
appropriate to model the fast decay of the magnetic field with the
distance to the source. To overcome these difficulties, and in part motivated
by the strategy introduced in~\citep{2011arXiv1112.2622L}, we introduce the
following phenomenological current to include both the ideal MHD and the
force-free limits,
\begin{eqnarray}\label{ohm_ideal_forcefree}
   J^i &=& q [ (1-H)\, v^i + H\, v^i_{d} ] 
\\
   &+& \frac{\sigma}{1 + \zeta^2 } 
  \left[{\cal E}^i 
  + \frac{\zeta^2}{B^2} \{ (E^k B_k) B^i + \chi (E^2-B^2) E^i\} \right]~,
\nonumber
\end{eqnarray}
where $H$ is a function which vanishes whereas the ideal MHD 
limit is valid, and tends to $1$ whereas the force-free limit
is more appropriate. The anisotropic ratio $\zeta$, which can be
reinterpreted from the definition $\xi^2 b^2 \equiv \zeta^2$, can be 
conveniently set to be a constant in the region where the
force-free limit is valid. 
The physical condition $B^2-E^2>0$ is enforced through a new current term
proportional to an anomalous conductivity $\chi$, which only appears
whenever $B^2<E^2$. Overdamping of the electric field is avoided
by setting this anomalous conductivity to the characteristic
decay time $\chi \approx (\alpha \sqrt{\gamma} \sigma \Delta t )^{-1}$,
which can be estimated from the time evolution of $B^2-E^2$.

Let us consider the particular astrophysical scenario of magnetized
neutron stars. The large fluid conductivity, both inside and outside
the star, is modeled by using a large constant $\sigma \approx 10^5$.
The anisotropic ratio, which defines the regions described either with
the ideal MHD or the force-free limits, is defined as $\zeta = H \sigma$.
This choice ensures that the interior of the star (i.e., $H=0$) is
dominated by a large isotropic conductivity, reducing the system of
equations to the ideal MHD limit. The exterior of the star (i.e., $H=1$)
is dominated by the anisotropic terms which enforce the force-free
condition. The kernel function $H$ is defined such that vanishes inside
the star and its value becomes unity outside. A smooth transition between
the inner and the outer region is achieved by using
\begin{equation}\label{eq:kernel}
    H(\rho \,,\rho_o)= \frac{2}{1 + e^{2\,K\,(\rho - \rho_o)}}
\end{equation}
We typically adopt $K \approx 0.001/\rho_{atm}$ and 
$\rho_o \approx 50-400\, \rho_{atm}$, being $\rho_{atm}$ 
the value for the density of the magnetosphere.

%%%%%%%%%%%%%%%%%%%%%%%%%%%%%%%%%%%%%%%%%%%%%%%%%%%%%%%%%%%%%%%%%%%%%%%%%%%%%%%
%%%%%%%%%%%%%%%%%%%%%%%%%%%%%%%%%%%%%%%%%%%%%%%%%%%%%%%%%%%%%%%%%%%%%%%%%%%%%%%
\section{Hyperbolic systems with relaxation terms}
\label{sec:hyperbolic_relaxation}
%%%%%%%%%%%%%%%%%%%%%%%%%%%%%%%%%%%%%%%%%%%%%%%%%%%%%%%%%%%%%%%%%%%%%%%%%%%%%%%
%%%%%%%%%%%%%%%%%%%%%%%%%%%%%%%%%%%%%%%%%%%%%%%%%%%%%%%%%%%%%%%%%%%%%%%%%%%%%%%

The general system of relativistic resistive MHD equations
(\ref{maxwell11}-\ref{fluid13},\ref{ohm_ideal_forcefree}) brings about a
delicate issue when the conductivity in the plasma undergoes very large spatial
variations. In regions with high conductivity, in fact, the system will
evolve on timescales which are very different from those in the low-conductivity
region. Mathematically, therefore, the problem can be regarded as a
hyperbolic system with relaxation terms which requires special
care to capture the dynamics in a stable and accurate manner.
The prototype of these systems can be written as
\begin{equation}\label{stiff_equation}
    \partial_t {\bf U} = F({\bf U}) + \frac{1}{\epsilon} R({\bf U})
\end{equation}
where $\epsilon >0$ is the relaxation time. In the limit
$\epsilon \rightarrow \infty$ the system is hyperbolic
with spectral radius $c_h$ (i.e., the absolute value of the maximum
eigenvalue). In the other limit $\epsilon \rightarrow 0$ the system
is clearly stiff since the time scale of the relaxation
(or stiff term) $R({\bf U})$ is much smaller than the maximum speed $c_h$
of the hyperbolic part $F({\bf U})$. 

In the stiff limit ($\epsilon \rightarrow 0$) the stability of an explicit
time evolution scheme is only achieved with a time step size
$\Delta t \leq \epsilon$, a much stronger restriction than the
CFL condition $\Delta t \leq \Delta x / c_h $ of the hyperbolic systems.
The development of stable and efficient numerical schemes to overcome
this restrictive constraint is challenging, since in many applications
the relaxation time can vary many orders of magnitude.

Different alternatives to deal with the inherent stiffness of the
relativistic resistive MHD equations has been proposed in the last decade;
combination of splitting methods and analytical solutions
~\citep{Komissarov2007,2010ApJ...716L.214Z,2011ApJ...735..113T},
discontinuous Galerkin methods~\citep{2011MNRAS.418.1004Z,2009JCoPh.228.6991D}
and Implicit-Explicit (IMEX) Runge-Kutta methods~\citep{2009MNRAS.394.1727P,
2012arXiv1205.2951B,2012arXiv1208.3487D}. The following subsections 
summarize the IMEX Runge-Kutta schemes, a family of time integrators
which are able to deal with the potentially stiffness issues and
are relatively easy to incorporate into an existing relativistic
ideal MHD code.

%%%%%%%%%%%%%%%%%%%%%%%%%%%%%%%%%%%%%%%%%%%%%%%
\subsection{Implicit-Explicit Runge-Kutta methods}
\label{subsec:imex}
%%%%%%%%%%%%%%%%%%%%%%%%%%%%%%%%%%%%%%%%%%%%%%%

An efficient way to solve the hyperbolic-relaxation systems
is based on the IMEX Runge-Kutta methods. Within this scheme,
all the fields are evolved by using a standard explicit time
integration except the potentially stiff terms,
which are evolved with an implicit time discretization.
For the generic system (\ref{stiff_equation}) this scheme
takes the form~\citep{ParRus:2005}
\begin{eqnarray}\label{IMEX}
   {\bf U}^{(i)} = {\bf U}^n &+& \Delta t \sum_{j=1}^{i-1} {\tilde{a}}_{ij} F({\bf U}^{(j)}) 
\nonumber \\
     &+& \Delta t  \sum_{j=1}^{\nu} a_{ij} \frac{1}{\epsilon} R({\bf U}^{(j)}) \\
  {\bf U}^{n+1} = {\bf U}^n &+& \Delta t \sum_{i=1}^{\nu} {\tilde{\omega}}_{i} F({\bf U}^{(i)})
     + \Delta t  \sum_{i=1}^{\nu} \omega_{i} \frac{1}{\epsilon} R({\bf U}^{(i)}) 
\nonumber
\end{eqnarray}
where ${\bf U}^{(i)}$ are the auxiliary intermediate values of the Runge-Kutta.
The coefficients can be represented as $\nu \times \nu$ matrices
$\tilde{A}= (\tilde{a}_{ij})$ and $A= (a_{ij})$ such that the resulting scheme
is explicit in $F$ (i.e.,$\tilde{a}_{ij}=0$ for $j \ge i$) and implicit in $R$.
An IMEX Runge-Kutta is characterized by these two matrices and the coefficient
vectors $\tilde{\omega}_i$ and $\omega_i$. Notice that at each substep 
the auxiliary intermediate values ${\bf U}^{(i)}$ involves solving 
an implicit equation. Since the simplicity and efficiency of solving the
implicit part at each step is of great importance, it is natural to consider
diagonally implicit Runge-Kutta (DIRK) schemes ($a_{ij}=0$ for $j > i$) for
the stiff terms. A deeper discussion on
the IMEX schemes and the detailed form of the schemes considered here
are presented in appendix~\ref{appendixA}.

%%%%%%%%%%%%%%%%%%%%%%%%%%%%%%%%%%%%%%%%%%%%%%%
\subsection{Solving generic systems with IMEX schemes}
\label{subsec:applying}
%%%%%%%%%%%%%%%%%%%%%%%%%%%%%%%%%%%%%%%%%%%%%%%

The vector of evolved fields $\bf{U}$ can be split in two sets
of variables $(\bf{V},\bf{W})$, depending on whether or not they contain any
relaxation term in their evolution equations. The evolution system can
then be generically written as
\begin{eqnarray}\label{split}
    \partial_t {\bf W} &=& F_W({\bf V},{\bf W}) \\
    \partial_t {\bf V} &=& F_V({\bf V},{\bf W}) 
        + \frac{1}{\epsilon} R_V({\bf V},{\bf W}) ~,
\end{eqnarray}
where we have considered that the relaxation parameter $\epsilon$ can be
any function not depending directly on the present value of
the ${\bf V}$-fields. The procedure to compute each auxiliary step
${\bf U}^{(i)}$ can be split in two stages:
\begin{enumerate}
  \item compute first the intermediate values $\{\bf{V^*},\bf{W^*}\}$ which involves
   information from previous steps, 
  \begin{eqnarray}\label{first_step}
   {\bf W}^{*} = {\bf W}^n &+& \Delta t~ \sum_{j=1}^{i-1}~ {\tilde{a}}_{ij} F_W({\bf U}^{(j)}) 
\nonumber \\
   {\bf V}^{*} = {\bf V}^n &+& \Delta t~ \sum_{j=1}^{i-1}~ {\tilde{a}}_{ij} F_V({\bf U}^{(j)}) 
\nonumber \\
     &+& \Delta t~ \sum_{j=1}^{i-1}~ a_{ij} \frac{1}{\epsilon^{(j)}} R_V({\bf U}^{(j)})  ~~.
   \end{eqnarray} 
  \item include the relaxation term at the present time
    by solving the implicit equation
   \begin{eqnarray}\label{second_step}
         {\bf W}^{(i)} &=& {\bf W}^{*} 
     \nonumber  \\
         {\bf V}^{(i)} &=& {\bf V^*} 
         + a_{ii}~\frac{\Delta t}{\epsilon^{(i)}}~R_V({\bf V}^{(i)},{\bf W}^{(i)})
   \end{eqnarray}
   which clearly involves only the ${\bf V}$-fields.
\end{enumerate}

The complexity of inverting this implicit equation depends on the
particular form of the relaxation terms. From now on we will
restrict ourselves to the algebraic case $R_V({\bf U})=f({\bf U})$.
Next it is described two different ways to solve this implicit
equation; the first one can only be applied when $R_V({\bf U})$
is a linear function, whereas the second one
allows $R_V({\bf U})$ to have any non-linear dependence.

%%%%%%%%%%%%%%%%%%%%%%%%%%%%%%%%%%%%%%%%%%%%%%%
\subsubsection{$R_V$ depending linearly on ${\bf V}$}
\label{subsubsec:linearR}
%%%%%%%%%%%%%%%%%%%%%%%%%%%%%%%%%%%%%%%%%%%%%%%

The simplest case, however enough to cover a broad range
of interesting situations, is to consider a linear relaxation term
\begin{equation}\label{stiff_part}
  R_V({\bf V},{\bf W}) = A({\bf W}) {\bf V} + S(\bf{W}) ~.
\end{equation}  
The implicit equation (\ref{second_step}) can then be trivially
solved
\begin{eqnarray}\label{invert_matrix}
    {\bf V}^{(i)} &=& M \left[ {\bf V^*} + a_{ii}~\frac{\Delta t}{\epsilon^{(i)}}~S({\bf W^{(i)}}) \right]
\nonumber  \\
     M &=& [I - a_{ii}~\frac{\Delta t} {\epsilon^{(i)}} A({\bf W^{(i)}})]^{-1} ~.
\end{eqnarray}  
The matrix inversion can be performed analytically and written in
a compact form for most of the interesting cases, so that the
implicit step can be solved in a completely explicit way.

%%%%%%%%%%%%%%%%%%%%%%%%%%%%%%%%%%%%%%%%%%%%%%%
\subsubsection{$R_V$ depending non-linearly on ${\bf V}$}
\label{subsubsec:nonlinearR}
%%%%%%%%%%%%%%%%%%%%%%%%%%%%%%%%%%%%%%%%%%%%%%%

In the more general case --with an arbitrary non-linear dependence--
it is usually not feasible to solve analytically the implicit step,
requiring some approximation to find the solution.
A convenient approach to solve this problem is to linearize
the stiff term around an approximate solution
$\{ {\bf \bar{V}} ,{\bf W}^{(i)}\}$, such that
\begin{eqnarray}\label{linearization2}
  R_V({\bf V}^{(i)},{\bf W}^{(i)}) &\approx& R_V({\bf \bar{V}},{\bf W}^{(i)}) 
\\
  &+& \left( \frac{\partial R_V}{\partial {\bf V}} \right)_{{\bf \bar{V}},{\bf W}^{(i)}}
                             ({\bf V}^{(i)} - {\bf \bar{V}})~.
\nonumber
\end{eqnarray}
Notice that we are linearizing around the solution ${\bf W}^{(i)}$, which
is already known at the beginning of the implicit step.

By defining $A \equiv \left( \frac{\partial R_V}{\partial {\bf V}}
\right)_{{\bf \bar{V}},{\bf W}^{(i)}}$, and substituting the previous
expansion (\ref{linearization2}) in (\ref{second_step}), it is obtained
\begin{equation}\label{second_step_linear}
  {\bf V}^{(i)} = {\bf V^*} + a_{ii}~\frac{\Delta t}{\epsilon^{(i)}}
    [ R_V({\bf \bar{V}}) + A ({\bf V}^{(i)} - {\bf \bar{V}}) ]
\end{equation}
This implicit equation can be written, after some manipulations, 
in the following way
\begin{eqnarray}\label{invert_matrix_linearized_ final}
    {\bf V}^{(i)} &=& {\bf \bar{V}} + M [ {\bf V}^* - {\bf \bar{V}} +
      a_{ii}~\frac{\Delta t} {\epsilon^{(i)}} ~R_V({\bf \bar{V}},{\bf W}^{(i)}) ]
\nonumber \\
M &\equiv& [I - a_{ii}~\frac{\Delta t} {\epsilon^{(i)}} A({\bf \bar{V}},{\bf W}^{(i)})]^{-1}
\end{eqnarray}  
The final expression (\ref{invert_matrix_linearized_ final}) 
can be solved through a Newton-Raphson iterative procedure such that,
at each iteration $m$, uses an initial guess ${\bf \bar{V}}={\bf V}^{(i)}_{(m-1)}$
to find the next approximate solution ${\bf V}^{(i)}_{(m)}$.

%%%%%%%%%%%%%%%%%%%%%%%%%%%%%%%%%%%%%%%%%%%%%%%%%%%%%%%%%%%%%%%%%%%%%%%%%%%%%%%
\section{Numerical evolution of the Resistive MagnetoHydroDynamics system}
\label{sec:rmhd}
%%%%%%%%%%%%%%%%%%%%%%%%%%%%%%%%%%%%%%%%%%%%%%%%%%%%%%%%%%%%%%%%%%%%%%%%%%%%%%%

We adopt finite difference techniques on a regular Cartesian grid to solve
the problems of interest. To ensure sufficient resolution is achieved in an efficient
manner we employ adaptive mesh refinement (AMR) via the HAD computational
infrastructure~\footnote{publicly available at http://had.liu.edu}
that provides distributed, Berger-Oliger style
AMR~\citep{Liebling} with full sub-cycling in time, together with
an improved treatment of artificial boundaries~\citep{Lehner:2005vc}.
The refinement regions are determined using truncation error estimation provided
by a shadow hierarchy~\citep{Pretoriusphd} which adapts dynamically to ensure
the estimated error is bounded within a pre-specified tolerance.
The spatial discretization of the geometry is performed using a fourth order
accurate scheme, while that High Resolution Shock Capturing methods based
on the HLLE flux formula with PPM reconstruction are used to discretize the
resistive MHD variables~\citep{Anderson:2006ay,Anderson:2007kz}.
The time-evolution is performed through the method of lines using
a third order accurate Implicit-Explicit Runge-Kutta integration scheme
described in the previous section. We adopt a Courant parameter
of $\lambda = 0.25$ so that $\Delta t_l = 0.25 \Delta x_l$ on each
refinement level $l$. On each level, one therefore ensures that the
Courant-Friedrichs-Levy~(CFL) condition
dictated by the principal part of the equations is satisfied.

%%%%%%%%%%%%%%%%%%%%%%%%%%%%%%%%%%%%%%%%%%%%%%%
\subsection{Evolution of the electric field}
\label{subsec:dtE}
%%%%%%%%%%%%%%%%%%%%%%%%%%%%%%%%%%%%%%%%%%%%%%%

The relaxation terms of the resistive MHD system are associated to
the current, which mainly appears in the time evolution equation of the
electric field. The evolved fields can then be split into
and non-stiff ${\bf W} = \{ D,\tau, S_i, B^i, \psi,\phi, q \}$
and potentially stiff ${\bf V}= \{ E^i\}$.
The evolution of the non-stiff fields is performed by the
explicit part of the IMEX Runge-Kutta, and it is very similar to a
standard implementation of the ideal MHD equations. The evolution
of the electric field contains in addition the relaxation terms, namely
\begin{eqnarray}\label{split_stiff_part}
    \partial_t (\sqrt{\gamma} {\bf E}) &=& F_E + (\sqrt{\gamma} R_E) ~~.
\\
  F_E &=& - \partial_k [\sqrt{\gamma} \left( -\beta^k E^i 
   - \alpha ( \epsilon^{ikj} B_j - \gamma^{ik} \psi ) \right) ] ~,
\nonumber \\
  & & - \sqrt{\gamma} E^k (\partial_k \beta^i) 
   + \sqrt{\gamma} \psi \left( \gamma^{ij} \partial_j \alpha
   - \alpha \gamma^{jk} \Gamma^i_{jk} \right)
\nonumber \\
  && - \alpha \sqrt{\gamma} J^i_{e} ~,
\nonumber\\ 
  R_E &=& -\alpha J^i_{s} ~.
\nonumber
\end{eqnarray}  
where the factor $1/\epsilon$, corresponding to the fluid conductivity,
is absorbed in the definition of $R_E$. The current has been split
into a potentially stiff part, $J^i_{s}$, and the terms which can be
treated explicitly, $J^i_{e}$. For the phenomenological Ohm's law
(\ref{ohm_ideal_forcefree}) these components can be written explicitly as
\begin{eqnarray}\label{ohm_ideal_forcefree_splitting}
   J^i_{e} &=& q [ (1-H)\, v^i + H\, v^i_{d} ] ~,
\\
   J^i_{s} &=& \frac{\sigma}{1 + \zeta^2 } 
  \left[{\cal E}^i 
  + \frac{\zeta^2}{B^2} \{ (E^k B_k) B^i + \chi (E^2-B^2) E^i\} \right]~.
\nonumber
\end{eqnarray}
Notice that, although the evolution of $q$ is driven by the current,
these terms do not become potentially stiff in this equation since 
they are not proportional to the field itself.
However, the delicate balance between the different fields in the current,
which allows to get finite values even for very high conductivities, may
be broken during the reconstruction of the fields at the interfaces.
These unacceptable large errors are prevented in the standard implementations
of the force-free equations by computing the charge density from the
constraint $q=\nabla_i E^i$ instead of using the charge conservation.
The resulting set of equations is still hyperbolic, since the charge
density only couples to the EM fields throughout the non-principal
term $q v^i$~\citep{2011CQGra..28m4007P}. Here we prefer
to keep the charge density as an evolution field and treat
all the fields in the same manner. The errors at the interfaces
are avoided by performing directly the reconstruction of the current
$J^i$, which is computed just after solving the stiff terms. This
ensures that the fluxes of $q$ will remain bounded between the values
given by well-defined neighboring points.

%%%%%%%%%%%%%%%%%%%%%%%%%%%%%%%%%%%%%%%%%%%%%%%
\subsection{Inversion from conserved to primitive variables}
\label{subsec:inversion}
%%%%%%%%%%%%%%%%%%%%%%%%%%%%%%%%%%%%%%%%%%%%%%%

The numerical evolution of the resistive MHD system (\ref{maxwell11}-\ref{fluid13})
involves the recovery, after each timestep, of the primitive fields
$\{ \rho,~ \epsilon,p,v^i,E^i,B^i,\psi,\phi,q\}$ from the conserved
or evolved fields $\sqrt{\gamma} \{ D,\tau, S_i, E^i, B^i, \psi,\phi,q \}$.
Although the conserved fields are just algebraic relations of 
the primitive ones, the opposite is not true; due to the enthalpy and
the Lorentz factor these quantities are related by complicated equations
that can only be solved numerically, except for particularly simple
equations of state.

The solution at time $t=(n+1)\Delta t$ is directly obtained, for most
of the conserved quantities, by evolving their (non-stiff) evolution
equations. However, the explicit evolution of the potentially stiff fields
only provides a partial solution. As explained in the previous
section, a complete solution for the electric field involves taking into 
account the relaxation terms by solving the corresponding implicit equation.
For a generic Ohm's law, these relaxation terms will depend on the
velocity and other primitive fields. Nevertheless, the recovery of the
primitive variables from the conserved ones involves all the fields,
including the electric field. This is a consistency constraint which implies
that the recovery process and the implicit step evolution must be solved
{\em at the same time}. We will next describe an iterative procedure to
evolve the stiff part and recover the primitive fields for the 
phenomenological current (\ref{ohm_ideal_forcefree}), as described in
subsection~\ref{subsubsec:nonlinearR}.

\begin{enumerate}
  \item To start the iterative process it is required an approximate solution
   --initial guess-- for the electric field ${\bar E}_i$ and the
   fluid unknowns of the system, that we have chosen to be the single combination
   $x \equiv h W^2$.
   The initial guess for this unknown is given simply by the previous time step
   ${\bar x} = x^{(n)}$. Possible choices for the electric field initial
   guess are: 
   \begin{itemize}
      \item the previous time step ${\bar E}_i = E^{(n)}_i$
      \item the ideal MHD limit ${\bar E}_i = - \epsilon_{ijk} v^j B^k$,
             which involves performing first the recovery in 
             the ideal MHD case (see appendix B for details).
      \item the approximate solution given by the explicit and previous
             implicit step evolutions ${\bar E}_i = E^*_i$.
      \item the trivial case ${\bar E}_i = 0$.
   \end{itemize}
   It may be difficult to estimate a priori which initial guess is more
   convenient. For this reason, our scheme starts  with the first option and,
   if no solution is found, tries sequentially the other choices.

    \item Subtract the electromagnetic contributions from the energy and
     momentum densities, 
     \begin{eqnarray}\label{con2prim_1}
       {\tilde \tau} &=& \tau - \frac{1}{2} (E^k E_k + B^k B_k) ~,~~~ \\
       {\tilde S}_{i} &=& S_{i} - \epsilon_{ijk} E^j B^k  
     \end{eqnarray}  
     such that the Lorentz factor can be computed as,
     \begin{equation}\label{con2prim_2}
        W^2 = \frac{x^2}{x^2 - {\tilde S}^i {\tilde S}_i} ~~~,~~
        c \equiv \frac{1}{W^2} = 1 - \frac{{\tilde S}^2}{x^2}
     \end{equation}

     \item Write also the pressure as a function of the conserved variables
      and the unknown $x$. For the ideal gas EOS $p = (\Gamma -1) \rho \epsilon$
      this relation is just
      \begin{equation}\label{c2p_pressure}
        p = \frac{\Gamma-1}{\Gamma} 
                     \left( \frac{x}{W^2} - \frac{D}{W} \right)
      \end{equation}

     \item  Obtain an equation $f(x)=0$, written in terms 
      of the unknown $x$ and the conserved fields, such that it 
      is satisfied only for true solutions of $x$. By using the previous
      expression (\ref{c2p_pressure}) in the definition of
      ${\tilde \tau}$, we can write
      \begin{equation}\label{fx_hybrid}
        f(x) = [ 1 - \frac{(\Gamma - 1)}{W^2 \Gamma}] x
              +[ \frac{\Gamma-1}{\Gamma W} -1] D        
             - {\tilde \tau} \,,
      \end{equation}
      where $W$ is computed through eq.(\ref{con2prim_2}). The equation
      $f(x)=0$ can be solved numerically by using an iterative Newton-Raphson solver.
      The solution in the iteration $m+1$ can be computed as
      \begin{equation}\label{newton-raphson}
         x_{(m+1)} = x_{(m)} - \frac{f(x_{(m)})}{f'(x_{(m)})} \,.
      \end{equation}   
      The derivative of the function $f(x)$ can be computed analytically,
      \begin{eqnarray}\label{derivativef}
        f'(x) &=& 1 - \frac{2 (\Gamma -1) {\tilde S}^2}{\Gamma x^2}
        \nonumber \\       
            &-& \frac{(\Gamma -1) c}{\Gamma} 
             + \frac{(\Gamma - 1) D {\tilde S}^2}{\sqrt{c} \Gamma x^3}
      \end{eqnarray}

      \item Update the primitive fields by using the relations
      \begin{eqnarray}\label{con2prim_3}
        v_i &=& \frac{{\tilde S}_i}{x} ~~,~~
        W^2 = \frac{x^2}{x^2 - {\tilde S}^2} ~~,~~
         h = \frac{x}{W^2} ~~~,~~~ 
        \nonumber \\
         p &=& \frac{\Gamma-1}{\Gamma} (h - \rho) ~~~,~~~
        \rho = \frac{D}{W} ~~~.
      \end{eqnarray}

      \item Update the electric field --with the updated values
      of the primitive fields-- by solving the implicit equation,
      corresponding to eq.~(\ref{second_step}),

      \begin{equation}
          E^i = E_{*}^i + a_{ii}~ {\Delta t} ~R_E^i  ~~,~~
      \end{equation}
       which can be formally solved with the method
       described in subsection~\ref{subsubsec:nonlinearR} for
      ${\bf V}^{(i)}=E^i$, that is,
      \begin{eqnarray}\label{invert_matrix_linearized5}
         E^i &=& {\bar E}^i 
          + M [ {E^*}^i - {\bar E}^i  + a_{ii}~ {\Delta t}~R^i_E ]
           \\
         M &=& [I - a_{ii} \Delta_t A]^{-1}~~,~~  
         A = \frac{\partial R_E^i}{\partial E^j} ~~.
      \end{eqnarray}
 
       For the phenomenological Ohm's law (\ref{ohm_ideal_forcefree}),
       the matrix $M$ to be inverted is
       \begin{eqnarray}\label{invert_matrix_linearized6}
        &M^{-1}& = \delta^i_j + {\tilde \sigma} \biggl[
                   W (\delta^i_j - v^i v_j) 
            \\
            &+& \frac{\zeta^2}{B^2} \left.\{  B^i B_j 
           + \chi [2 E^i E_j + \delta^i_j (E^2 - B^2)] \right.\} 
             \biggr]
           \nonumber
       \end{eqnarray}     
       with  ${\tilde \sigma}
       \equiv a_{ii}\, {\Delta t}\, \alpha\, \sigma/(1 + \zeta^2)$.
 
    \item Iterate until the solution $\{x,E^i\}$ satisfies
      their constitutive equations $f(x),f(E^i) \le 10^{-10}$,
      being $f(E^i)$ defined by equation (\ref{invert_matrix_linearized5}).

\end{enumerate}

In occasions the recovery procedure is unable to find a physical
state for a given set of conserved variables. In such cases, which
usually occur near a star's surface, failures can be avoided
by assuming that the fluid is isentropic in that timestep and 
therefore satisfying a polytropic EoS $p=K \rho^{\Gamma}$. Since 
the internal energy is also a function of the density 
(i.e., $\rho \epsilon = p/(\Gamma-1)$) for isentropic processes,
the conserved quantities are overdetermined and the energy equation
can be neglected in the recovery procedure, leading to a
more robust algorithm.

Notice also that, although our discussion was focused on the
phenomenological the Ohm's law (\ref{ohm_ideal_forcefree}),
the method described in subsection~\ref{subsubsec:nonlinearR} can be
applied to any algebraic form of the current. Even more general
cases with derivative terms can be considered, with the condition
that those must be evaluated at earlier times. In a similar way, the
method for linear relaxation terms described in 
subsection~\ref{subsubsec:linearR} can be generically used
for non-linear algebraic currents with the condition that the
non-linear terms are evaluated at previous time steps, as it
was considered in~\citep{2012ApJ...754...36A}.
This option does not require an initial guess for the electric field
and therefore may be more effective in avoiding unphysical states.

%%%%%%%%%%%%%%%%%%%%%%%%%%%%%%%%%%%%%%%%%%%%%%%%%%%%%%%%%%%%%%%%%%%%%%%%%%%%%%%
%%%%%%%%%%%%%%%%%%%%%%%%%%%%%%%%%%%%%%%%%%%%%%%%%%%%%%%%%%%%%%%%%%%%%%%%%%%%%%%
\section{Numerical simulations}
\label{sec:simulations}
%%%%%%%%%%%%%%%%%%%%%%%%%%%%%%%%%%%%%%%%%%%%%%%%%%%%%%%%%%%%%%%%%%%%%%%%%%%%%%%
%%%%%%%%%%%%%%%%%%%%%%%%%%%%%%%%%%%%%%%%%%%%%%%%%%%%%%%%%%%%%%%%%%%%%%%%%%%%%%%

In this section we report our numerical studies of astrophysical
scenarios involving the dynamical evolution of a rotating magnetized star
and its magnetosphere. The initial data of rigidly rotating neutron stars
is provided by the LORENE package
{\it Magstar}~\footnote{publicly available at http://www.lorene.obspm.fr},
which adopts a polytropic equation of state $P=K \rho^{\Gamma}$ 
with $\Gamma=2$, rescaled to $K=100$.
Because the fluid pressure in a neutron star is many orders of magnitude
larger than the electromagnetic one, moderate magnetic fields will have
an insignificant effect on both the geometry and the fluid structure,
and so they can be specified freely. For this reason we 
have chosen an initial poloidal magnetic field inside the star that
becomes dipolar in the external region. The electric fields are set by
assuming the ideal MHD condition, with an initial zero fluid velocity
in the magnetosphere.

During the evolution, which is performed with the methods described
in the previous sections, the ideal MHD and the force-free limits are
enforced inside/outside the star by using the phenomenological current
(\ref{ohm_ideal_forcefree}). We monitor the electromagnetic luminosity,
constructed from the Newman-Penrose scalar $\Phi_2$~\citep{Newman:1961qr},
\begin{equation}\label{L_em} 
  L_{\rm em} = \frac{{dE}^{\rm em}}{dt} = 
           \lim_{r \rightarrow \infty}  \int r^2 |\Phi_2|^2 d\Omega ~.
\end{equation}
that accounts for the energy carried off by outgoing waves to infinity
and it is equivalent to the Poynting luminosity at large distances.
Additionally we monitor the ratio of particular components
of the Maxwell tensor $\Omega_F = F_{tr}/F_{r\phi}$
which, in the stationary, axisymmetric case, can be interpreted as the
rotation frequency of the electromagnetic field~\citep{Blandford1977}.

%%%%%%%%%%%%%%%%%%%%%%%%%%%%%%%%%%%%%%%%%%%%%%%
\subsection{The aligned rotator}
\label{subsec:aligned_rotator}
%%%%%%%%%%%%%%%%%%%%%%%%%%%%%%%%%%%%%%%%%%%%%%%

%---------------------------------------------
\begin{figure}
 \begin{center}
\includegraphics[width = 70mm]{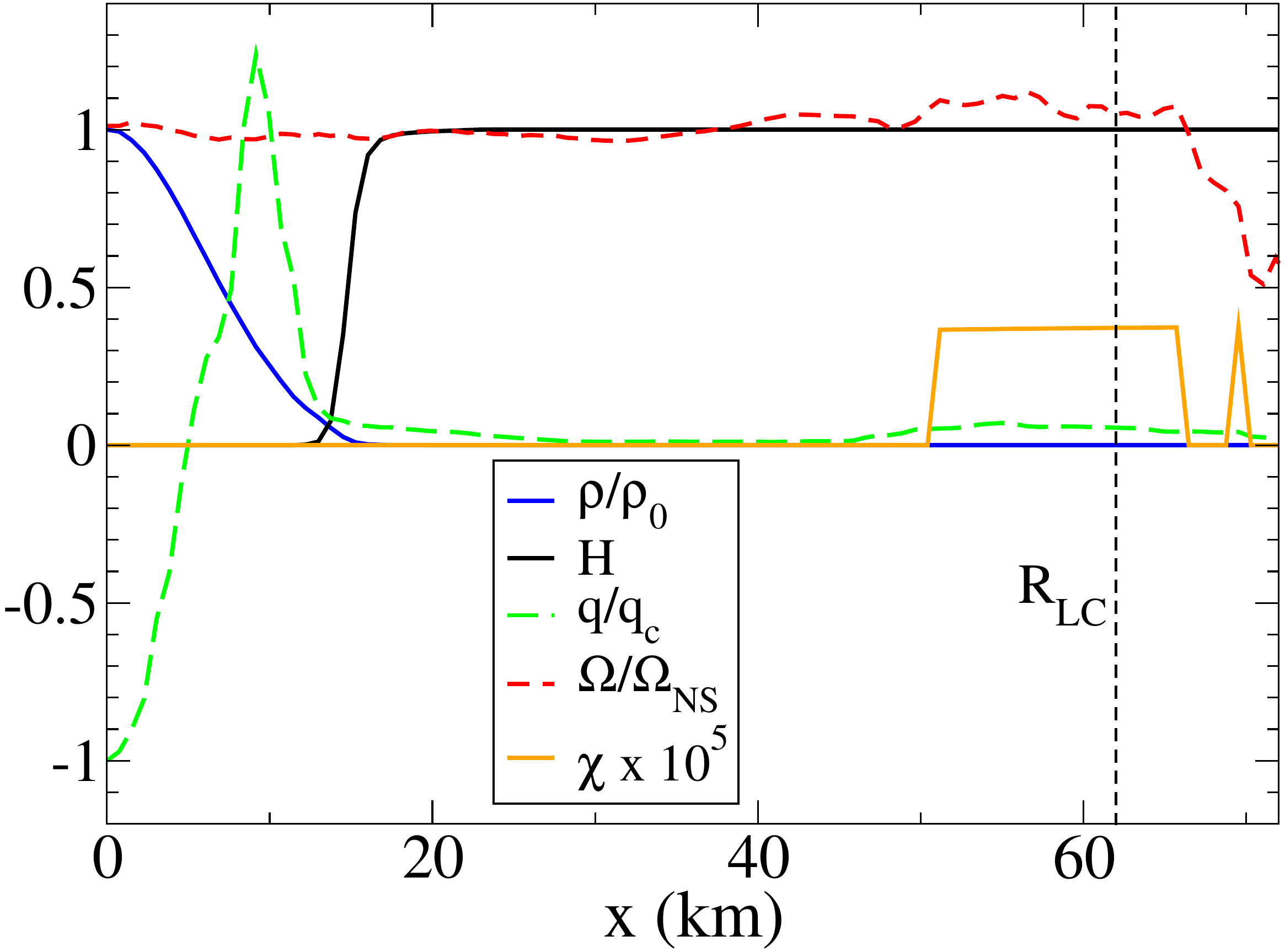}
 \end{center}
\caption{{\it Aligned rotator}. Several quantities displayed in the
equatorial plane as a function of the cylindrical radius after
two rotational periods. The kernel function $H$ indicates the value
of the density at which the current changes abruptly. The EM quantities
do not display any significant discontinuity in that region, as can be
appreciated for instance in the charge density. The magnetic fields
in the magnetosphere, up to the light cylinder, co-rotates with the frequency
of the star $\Omega_{NS}$. The anomalous resistivity appears only in
the regions with $E>B$, close and beyond the light cylinder.
}
\label{fig:pulsar1d}
\end{figure}
%
%---------------------------------------------

We consider first the evolution of an uniformly rotating stable star
of mass $M=1.58 M_{\odot}$ and equatorial/polar radius 
$R=16.1/10.6~{\rm km}$.
The star rotates with a period $T=1.3 {\rm ms}$, so that the
light cylinder is located at
$R_{\rm LC} = c/\Omega_{NS} = 62~{\rm km}$.
The strength of the magnetic field at the pole is 
$B_p = 1.8 \times 10^{14} G$. 
The numerical domain extends up to $L=300~{\rm km}$ and contains four
centered FMR grids with decreasing sizes (and twice better resolved)
such that the highest resolution grid has $\Delta x = 0.76~{\rm km}$
and extends up to $76~{\rm km}$ (i.e., beyond the light cylinder).

\begin{figure*}
%  \begin{center}
  \includegraphics[width = 70mm]{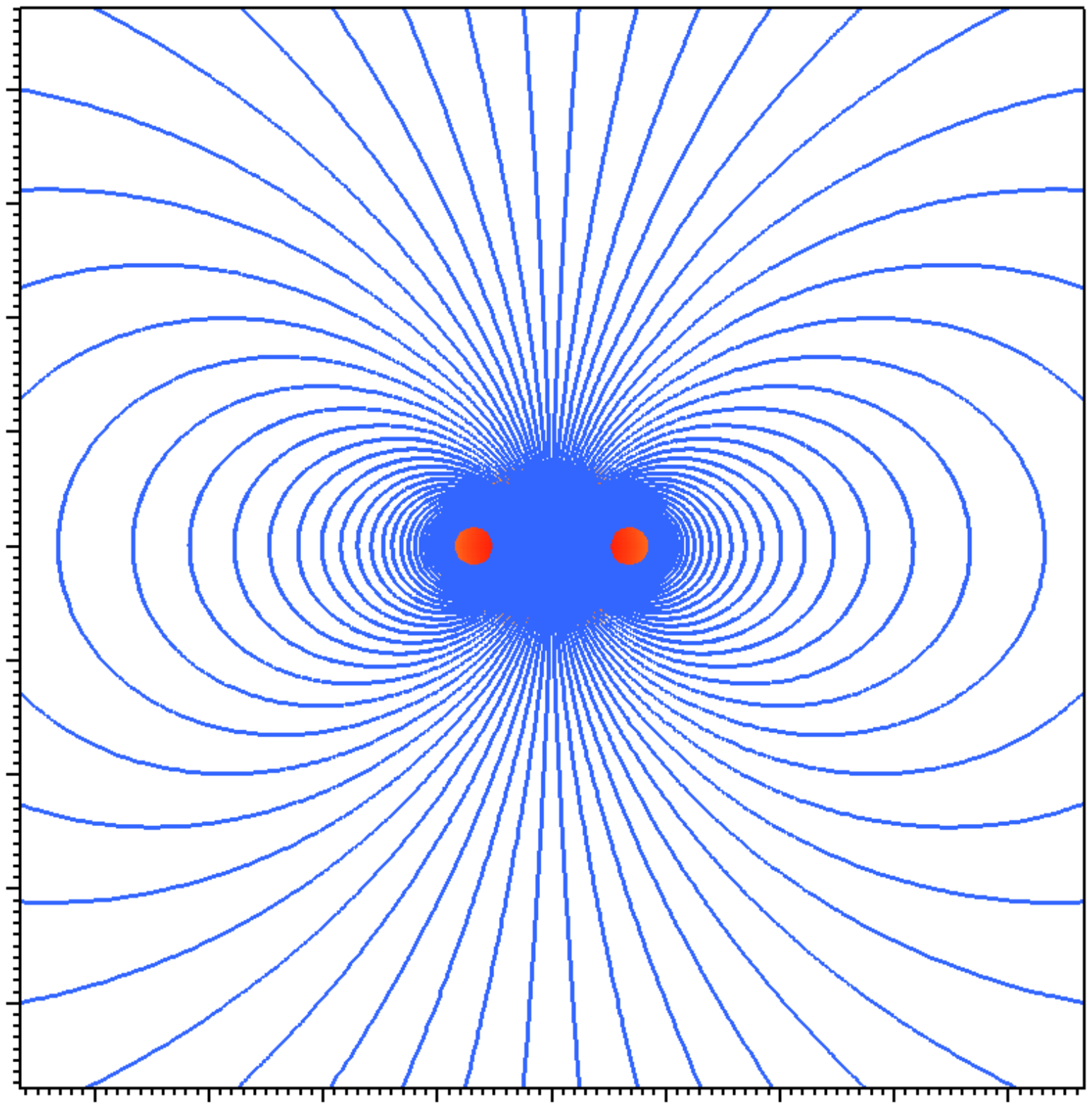} 
%  \hskip 1.0cm
  \includegraphics[width = 70mm]{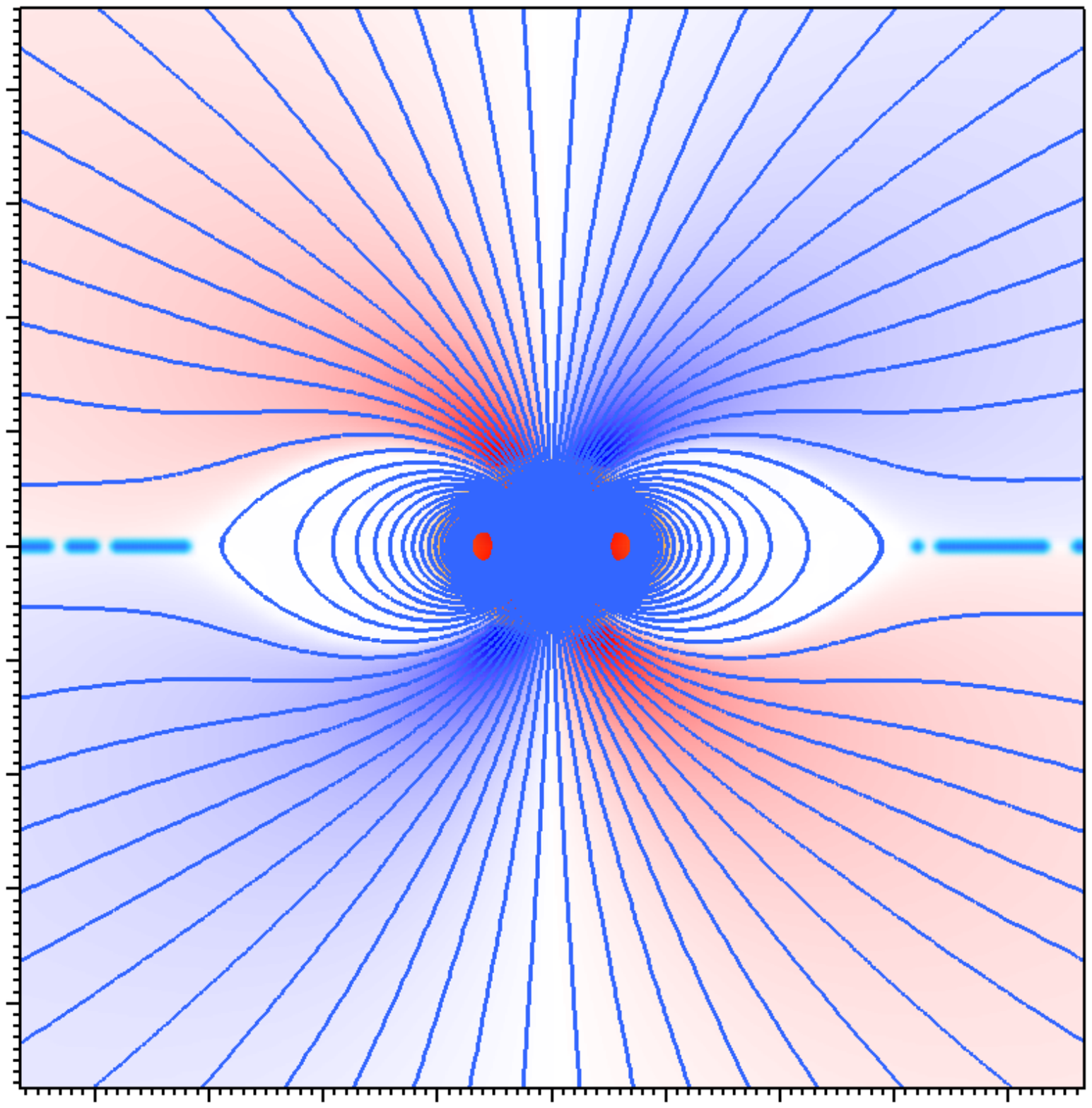}
%  \end{center}
\caption{{\it Aligned rotator}. The fluid density, the magnetic
field --poloidal components in lines and toroidal one in colors--
and the coefficient $\chi$ of the anomalous conductivity
on the $x=0$ plane at $t=0$ and after two rotational
periods of the star. The relaxed solution exhibits the known
properties of the aligned rotator solution, namely an opening of the
magnetic field lines roughly at the expected position 
$R_{\rm LC} \approx 4.0\, R_s$. These plots do not show the entire
computational domain.
}
\label{fig:pulsar_bfield}
\end{figure*}

This initial configuration is evolved until that the solution relaxes
to a quasi-stationary state. Different quantities are plotted
along the equatorial plane in fig.~\ref{fig:pulsar1d} and that
both the initial and the final magnetic field
solutions are displayed in fig.~\ref{fig:pulsar_bfield}.
The relaxed final state has the characteristic features observed
in previous works.
The magnetic fields are being dragged by the fluid rotation in the
interior of the star (i.e., as in the 
initial state), producing a tension that forces the magnetic fields
in the magnetosphere to co-rotate with the star up to the light
cylinder. Beyond this surface, the magnetic field lines open up,
creating a current sheet in the equatorial plane where the anomalous
resistivity in the current (or bringing back the neglected
fluid inertia) is necessary to preserve the physical condition $B^2>E^2$. 

%---------------------------------------------
%---------------------------------------------

We have computed the Poynting-vector luminosity at two
surfaces at $R_{ext}=\{76,114\}~{\rm km}$ located outside the light cylinder,
where the measures converge to a unique well-defined value.
The EM radiation is mainly dipolar
(i.e., around $90\%$ of the energy), with a small fraction in higher multipoles.
The luminosity can be compared with previous results in flat spacetime
geometry where the spherical star is modeled through inner boundary
conditions~\citep{2006ApJ...643.1139C,2006ApJ...648L..51S}
\begin{equation}\label{eq:lsd_dipole}
L_{\rm sd} = {1\over 4}B_{\rm pole}^2 R_{\rm NS}^2 c 
             \left({\Omega_{\rm NS} R_{\rm NS}\over c}\right)^4~.
\end{equation}
Our results agrees within a difference of $\approx 20\%$, where 
we have used $R_{\rm NS}=R_{eq}$. It is unclear where this small
disagreement may come from, since there are several possible
explanations; the ambiguity in the definition of the radius of
oblated stars, an excess of dissipation in the current sheet,
or purely strong gravitational effects, which may become important
due to the high compactness $M/R = 0.125$ of the star.

\begin{figure*}
\centerline{
\includegraphics[width = 75mm]{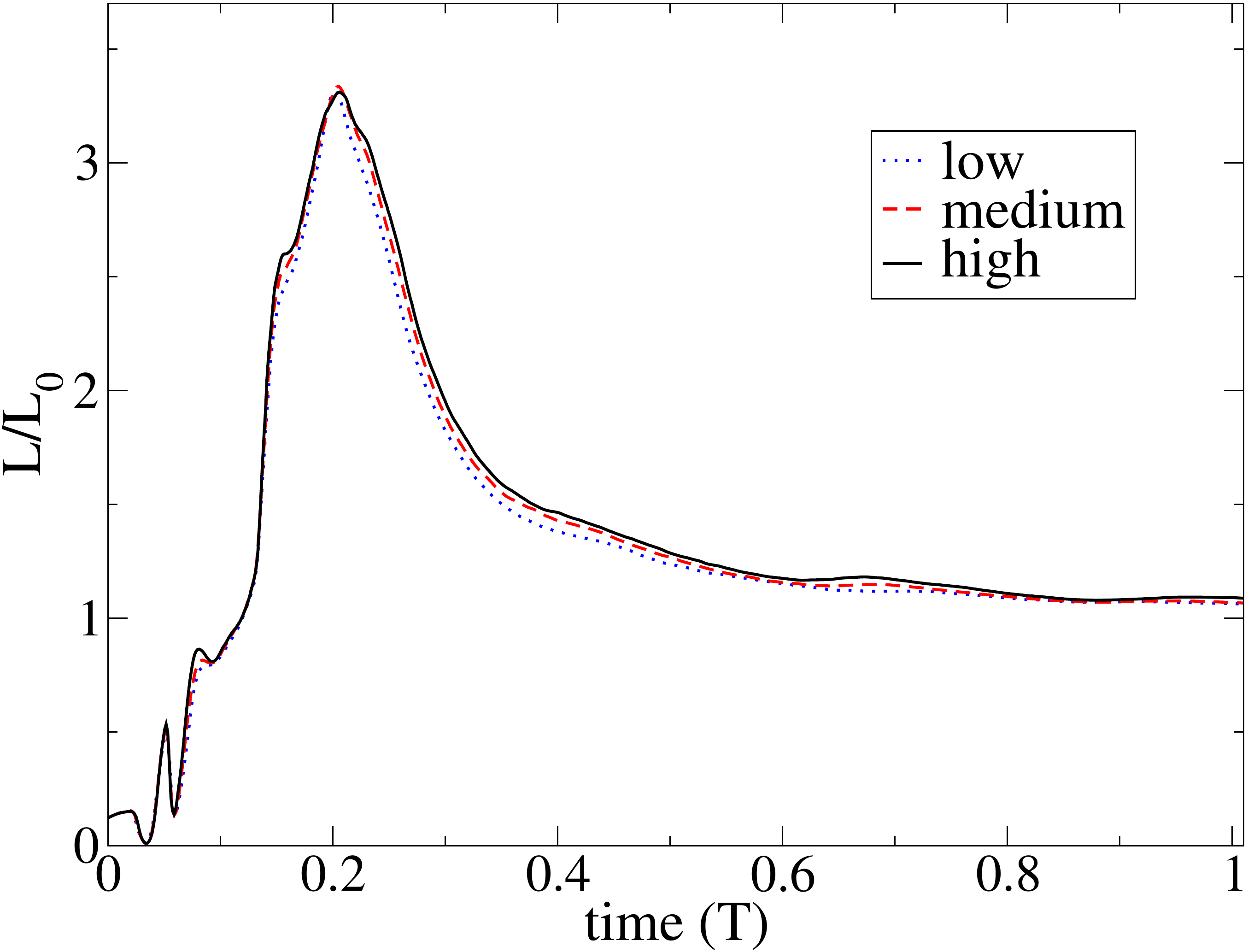} 
}
\caption{{\it Aligned rotator}. The EM
luminosity as a function of the rotational period
for three different resolutions 
$\Delta x = \{0.76,0.64,0.55\}{\rm km}$, showing the initial
transient followed by a fast decay to the quasi-stationary solution. 
The luminosity has been normalized with respect to the
asymptotic value, reached approximately after
$2$ rotational periods, of the low resolution simulation.
}
\label{fig:convergence_luminosity}
\end{figure*}

We have also monitored both the energy-momentum constraints
and the divergence constraints, checking that they remain small
and under control during the evolution. In particular,
$|\nabla \cdot B|/|B| \leq 0.05$ in all the domain but the current
sheet. By comparing the solutions obtained with three different
resolutions, each one improving a factor $1.18$ the previous space
discretization $\Delta x$, we have observed that the code
converges at $1.8$-order. The luminosity for these three
resolutions displayed in fig.~\ref{fig:convergence_luminosity}
shows that, in spite of the spasmodic reconnections happening in
the current sheet, the system converges to a quasi-stationary
solution with a steady luminosity.

%---------------------------------------------
%---------------------------------------------

%%%%%%%%%%%%%%%%%%%%%%%%%%%%%%%%%%%%%%%%%%%%%%%
\subsection{Collapse of a magnetized rotating neutron star}
\label{subsec:rotating_collapse}
%%%%%%%%%%%%%%%%%%%%%%%%%%%%%%%%%%%%%%%%%%%%%%%

%---------------------------------------------
\begin{figure*}
%\centerline{
\includegraphics[width = 56mm]{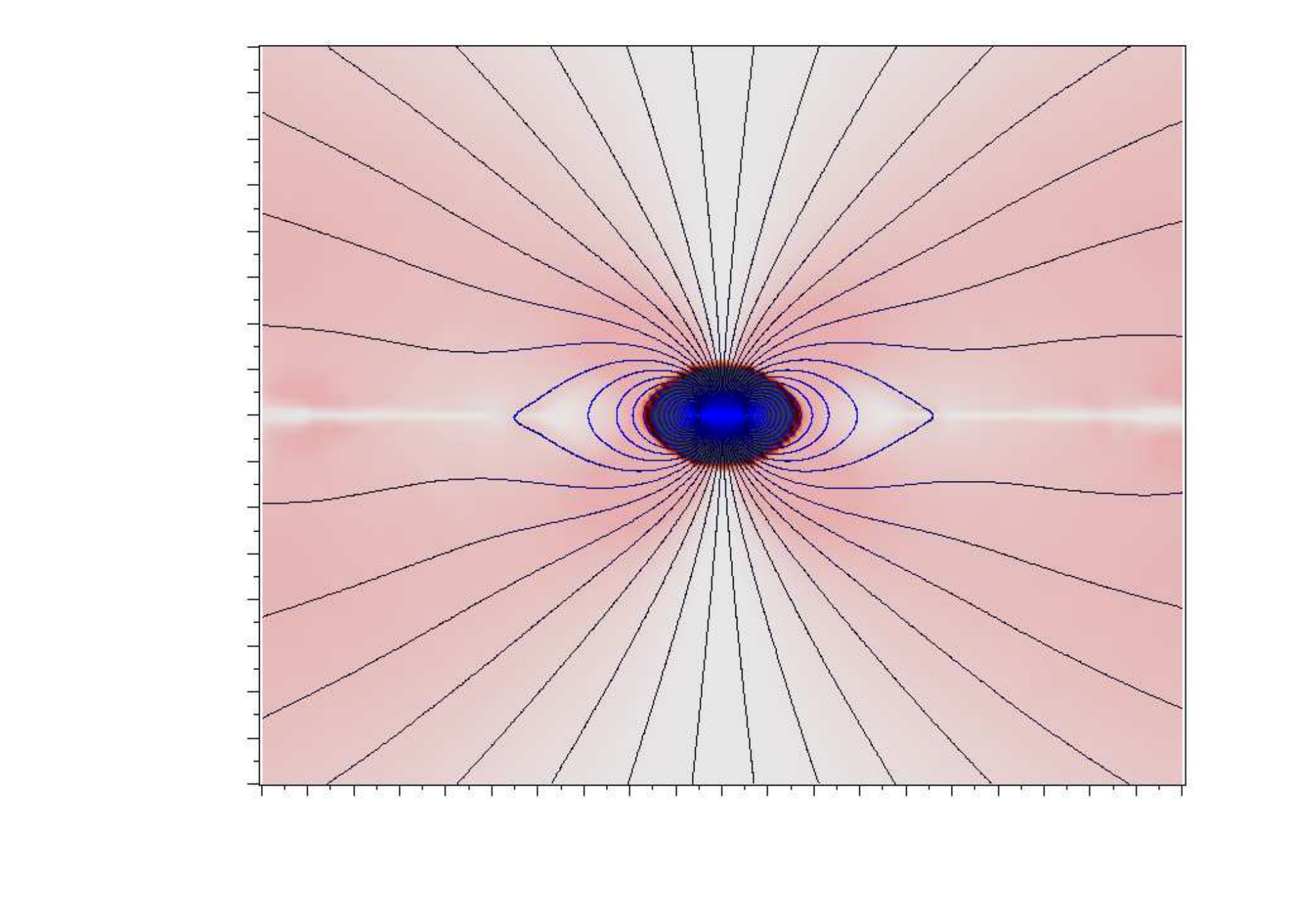}
\includegraphics[width = 56mm]{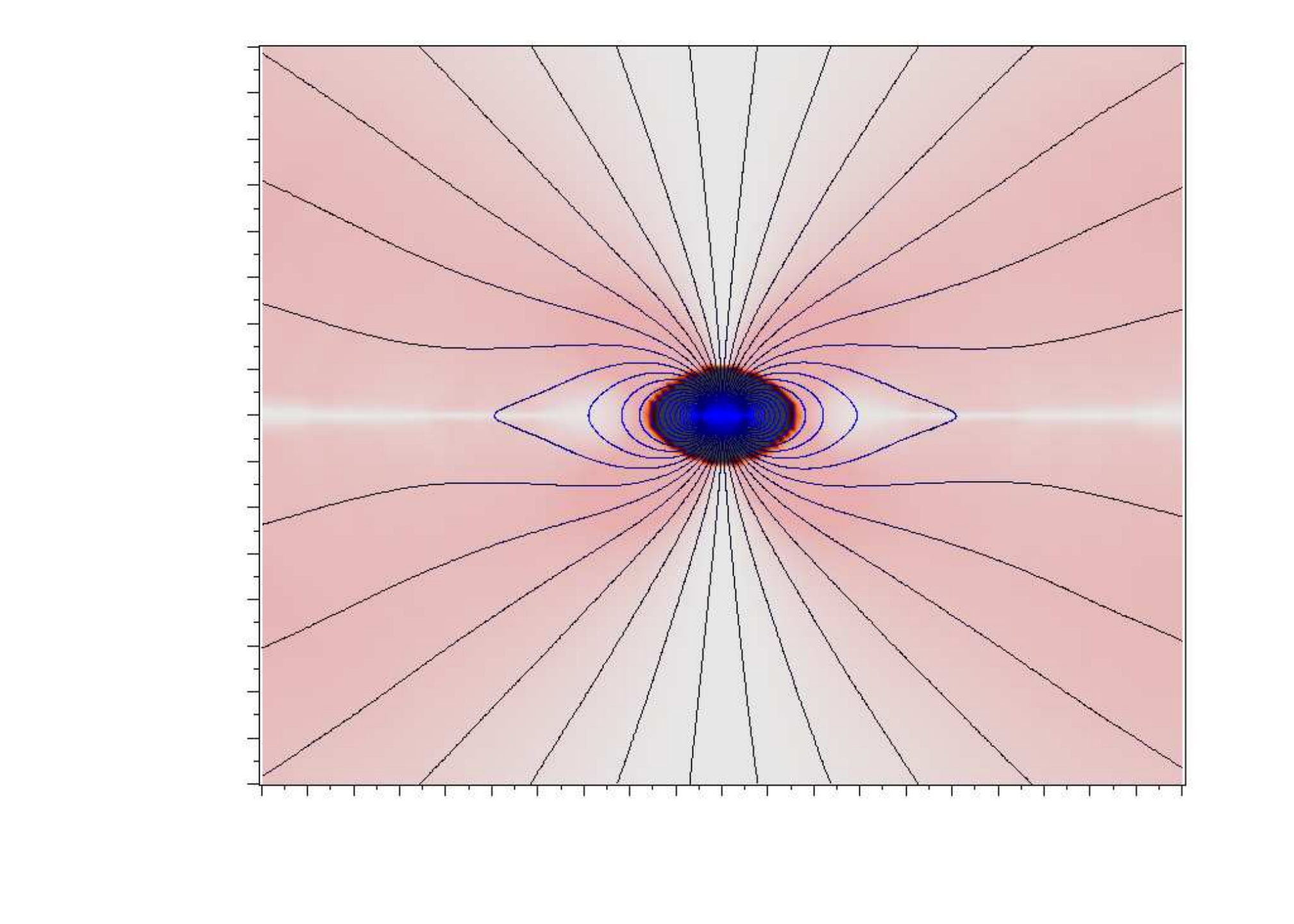} 
\includegraphics[width = 56mm]{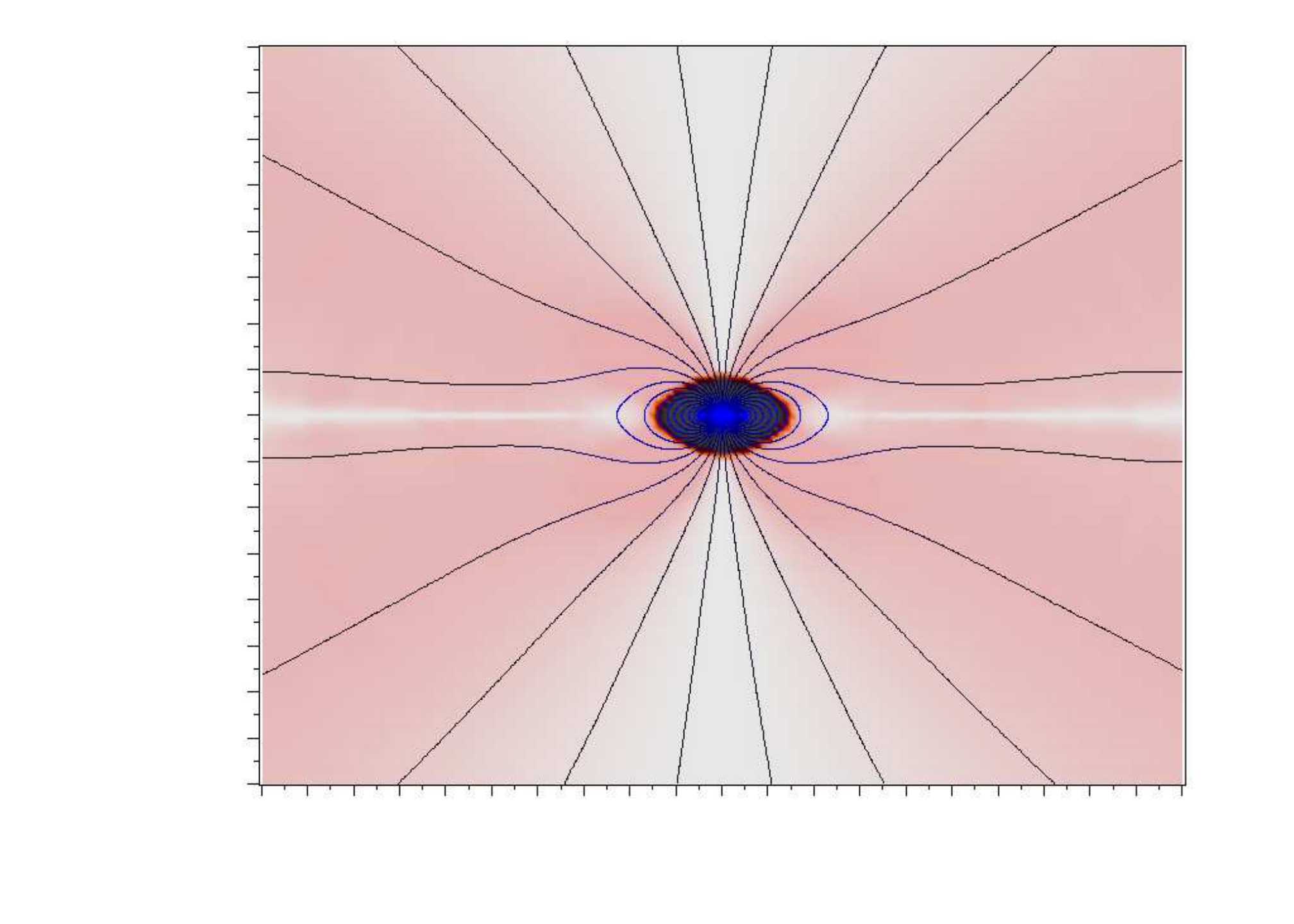}
\\
\includegraphics[width = 56mm]{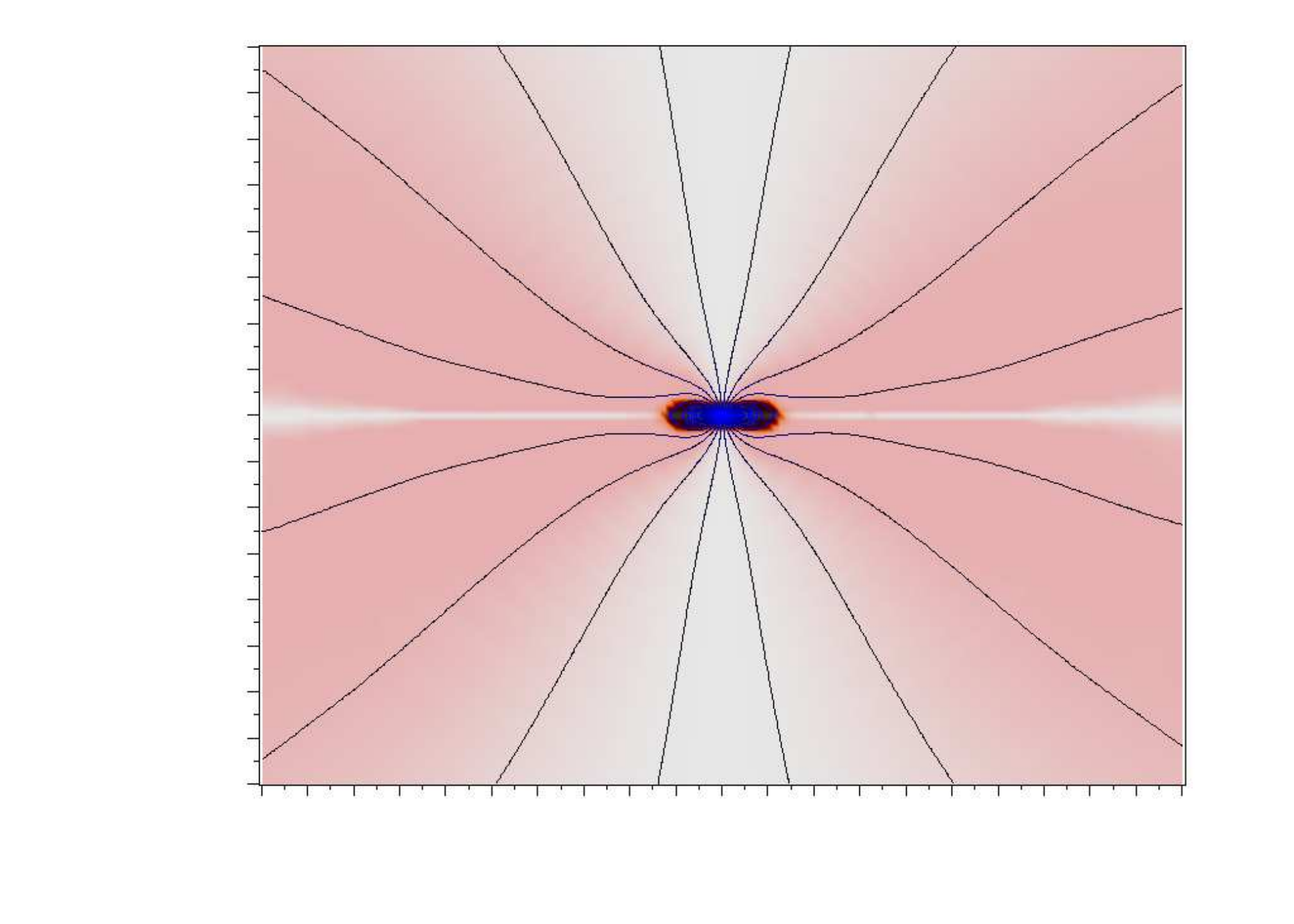} 
\includegraphics[width = 56mm]{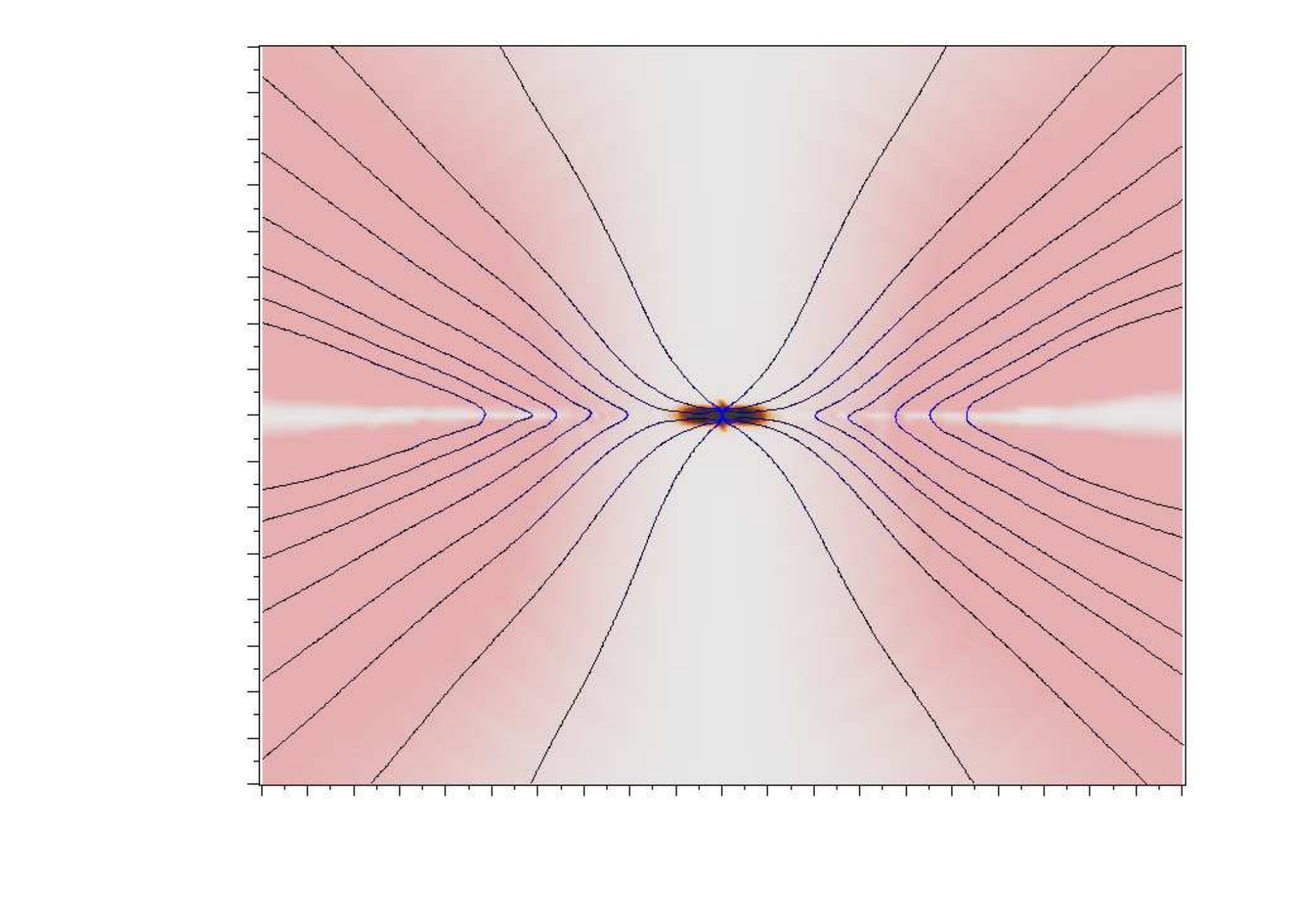}
\includegraphics[width = 56mm]{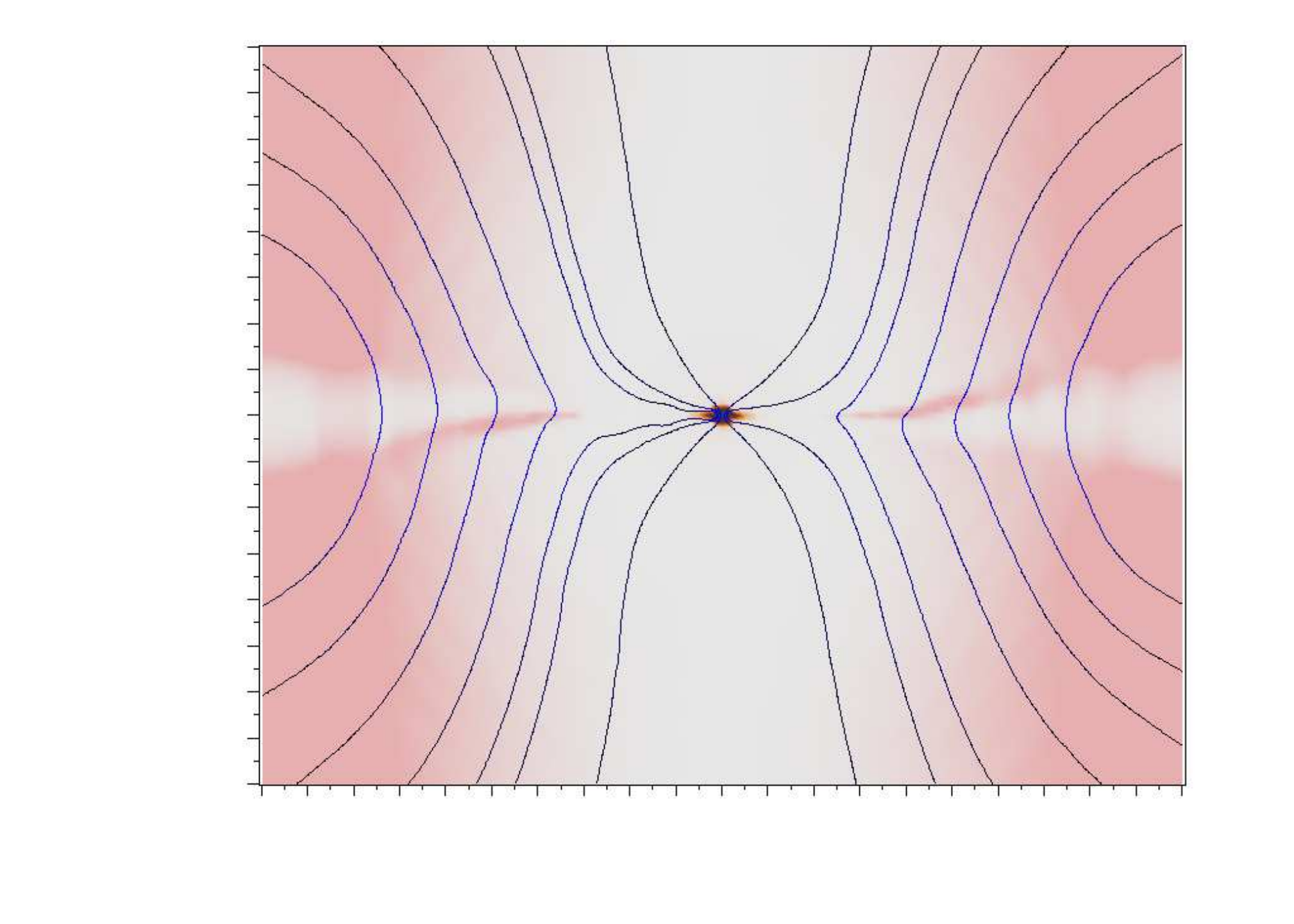} 
%}
\caption{{\it Collapse of a magnetized rotating star}. The fluid density,
the Poynting flux density and the poloidal magnetic field in the $x=0$
plane representative stages of the collapse corresponding to
$t=\{-0.43,-0.30,-0.18,-0.05,0.08,0.20\}$ ms. 
}
\label{fig:collapse_bfield}
\end{figure*}
%
%---------------------------------------------

After assessing the validity of our implementation with the aligned rotator
solution, we can consider a more challenging and dynamical case; the
collapse of an uniformly rotating magnetized neutron star to a black hole.
The initial data is the same as it was considered
in~\citep{2011arXiv1112.2622L}; a star lying on the unstable branch with
mass $M=1.84 M_{\odot}$ and equatorial/polar radius $R=10.6/7.3~{\rm km}$, 
rotating with a period $T=0.78 {\rm ms}$ so that the light cylinder is
located at $R_{\rm LC} = 37~{\rm km}$. 
The strength of the magnetic field at the pole is chosen to
be $B_p = 1.8 \times 10^{11} G$, although the results may be rescaled
to any strength as long as the magnetic pressure is much smaller than
the fluid one.
The numerical domain extends up to $L=300~{\rm km}$ and contains $6$
centered FMR grids with decreasing sizes such that the highest resolution
grid has $\Delta x = 0.19~{\rm km}$ and extends up to $21~{\rm km}$, while
that the second highest extends up to $44~{\rm km}$, beyond the initial
location of the light cylinder.

%---------------------------------------------
\begin{figure*}
\centerline{
\includegraphics[width = 75mm]{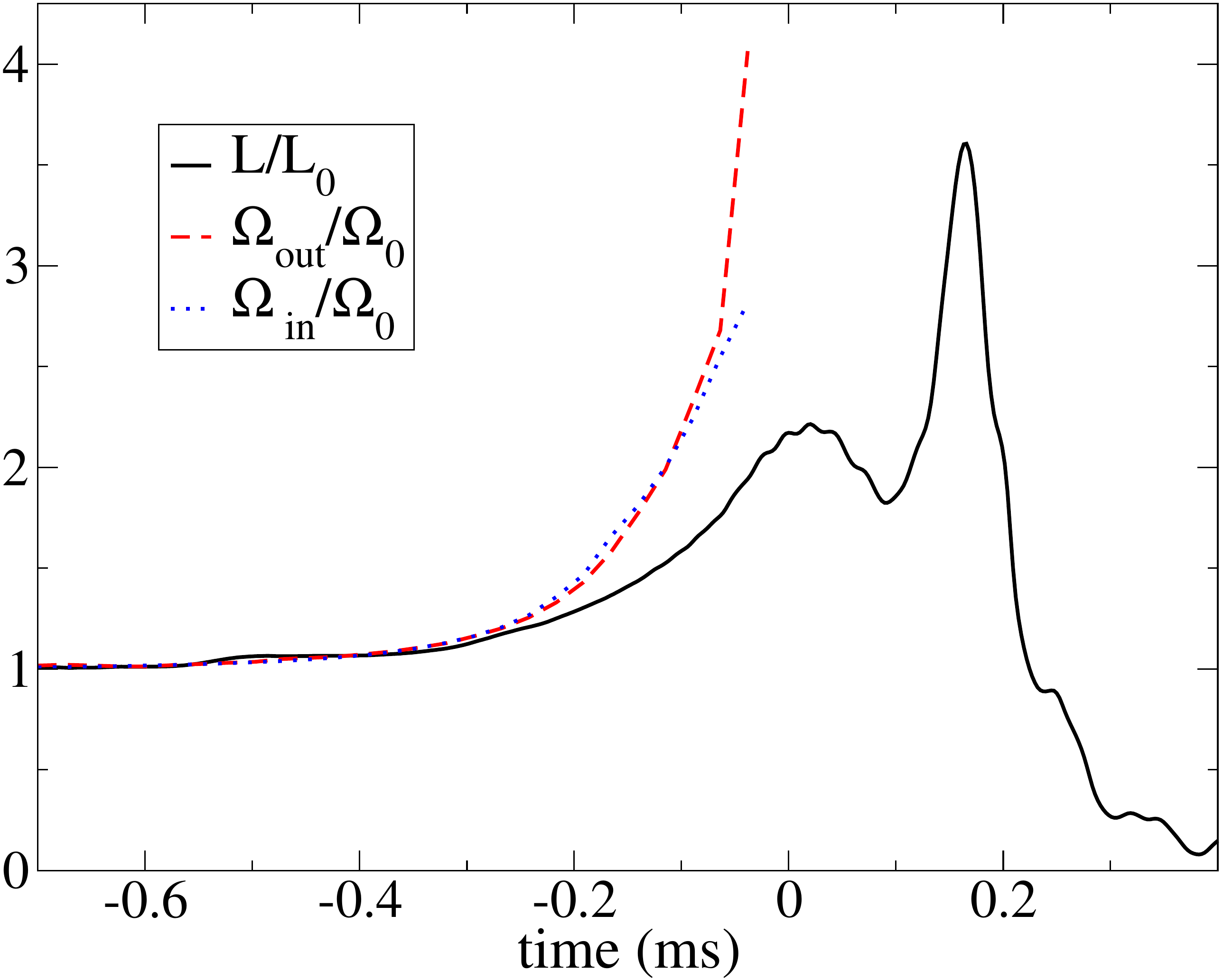} 
}
\caption{{\it Collapse of a magnetized rotating star}. The EM
luminosity and the angular frequency of the magnetic field
-- computed inside and outside the star-- as a function of time.
These quantities have been
normalized with respect to some reference values, calculated
once the system relaxes to a quasi-steady state (i.e., the aligned
rotator solution) at early times. The agreement between the
interior and the exterior angular velocity shows that it is
being propagated correctly through the surface of the star.
}
\label{fig:collapse_luminosity}
\end{figure*}
%---------------------------------------------

Small perturbations arising from numerical truncation errors are enough
to trigger the collapse of the unstable star. The horizon appears after
around $1 ms$, although the most dynamical part only stands
for the last $0.1 ms$, ending when all the matter disappears beyond the
horizon and the nearby magnetic fields reconnects in the
equatorial plane and escapes to infinity.
The conservation of angular momentum implies that
the angular velocity of the star increases
during the collapse, dragging the
magnetic field lines in the magnetosphere and bringing the light
cylinder closer to the star. The magnetic fields also grow
due to the magnetic flux conservation.
Once all the fluid has accreted onto the black hole, the magnetic
fields looses their anchorage, reconnects and propagates away
from the source. A significant fraction of the energy stored in the
magnetosphere is radiated to infinity in this burst.
The density of the star, the Poynting vector density $|\Phi_2|^2$ and
the magnetic fields are displayed at some representative stages of the
collapse in fig.~\ref{fig:collapse_bfield}.

The growth of the angular velocity and the magnetic field implies
that the luminosity of the aligned rotator~(\ref{eq:lsd_dipole})
during a quasi-adiabatic collapse will increase as
~$L_0 (R_{\rm NS}/R)^6$~\citep{2011PhRvD..83l4035L},
being $L_0$ the initial luminosity of the star.
However, since the collapse time is shorter than the star's period,
the outer part of the magnetosphere is not able to respond to the changes
in the start's surface, reducing the power of the
luminosity to $(R_{\rm NS}/R)^4$~\citep{2011arXiv1112.2622L}.
In addition, strong gravitational effects will soften the growth
of both the angular frequency and the radial magnetic field, 
leading to a much more moderate luminosity growth.

We have computed the electromagnetic luminosity in a sphere
located at $R_{ext}=76~{\rm km}$, beyond the light cylinder.
The EM radiation is mainly dipolar and grows during the
collapse, with a strong burst due to the reconnection when the
fluid is completely swallowed by the black hole. The luminosity and 
the angular velocity -- computed inside and outside the star-- 
are displayed in figure~\ref{fig:collapse_luminosity}.
The energy in the magnetosphere increases by a factor 
$C_{peak} \approx 2$ during the collapse. The total radiated energy can be
expressed as a fraction $\epsilon_{\rm rad}$ of the peak energy 
$C_{peak} E_{dipole,0}$, namely
\begin{equation}\label{energy_peak_magnetosphere}
   E_{\rm rad} \approx
   1.4\times 10^{47} \, C_{\rm peak}\, \epsilon_{\rm rad} 
   \left( \frac{B_{p}}{10^{15} G} \right)^2 \, {\rm erg}.
\end{equation}
where we have used 
$E_{\rm dipole,0} = 1.4 \times 10^{47} B_{\rm pole,15}^2 \, {\rm erg}$
for a star of radius $R_{NS} \approx 12~{\rm km}$
~\citep{2011arXiv1112.2622L}. In our simulation we have found
$\epsilon_{\rm rad} = 0.6$, implying that the system radiates 
$E_{\rm rad} \approx 1.6 \times 10^{47}$ergs during the collapse
(for a magnetic field of $10^{15}$G).
Notice that this value is different from the analytical estimates and indicates
the importance of the fast dynamic and strong gravitational effects in
this scenario.

%%%%%%%%%%%%%%%%%%%%%%%%%%%%%%%%%%%%%%%%%%%%%%%%%%%%%%%%%%%%%%%%%%%%%%%%%%%%%%%
%%%%%%%%%%%%%%%%%%%%%%%%%%%%%%%%%%%%%%%%%%%%%%%%%%%%%%%%%%%%%%%%%%%%%%%%%%%%%%%
\section{Summary}
\label{sec:discussion}
%%%%%%%%%%%%%%%%%%%%%%%%%%%%%%%%%%%%%%%%%%%%%%%%%%%%%%%%%%%%%%%%%%%%%%%%%%%%%%%
%%%%%%%%%%%%%%%%%%%%%%%%%%%%%%%%%%%%%%%%%%%%%%%%%%%%%%%%%%%%%%%%%%%%%%%%%%%%%%%

We have presented a formulation of the general relativistic
resistive MHD equations.
We have discussed different generalizations of the isotropic Ohm's law,
and constructed a phenomenological current such that
the system reduces either to the ideal MHD limit or to the force-free
approximation just by changing the ratio of isotropic/anisotropic
conductivities. We have explained how to deal with the potential 
stiffness of the equations by using the implicit-explicit
Runge-Kutta methods, showing how to perform the implicit evolution
of the electric field and the recovery of the primitive from the
conserved fields at the same time for any algebraic Ohm's law.
We implemented the formulation within the HAD computational 
infrastructure and revisited two interesting astrophysical
problems; the aligned rotator and the collapse of a rotating neutron star
to a black hole. None of these cases has a known analytical 
solution, although the first case has been studied extensively. 
We find a reasonable agreement between our results and previous studies
of the aligned rotator, recovering the same qualitative features and
approximately the same electromagnetic luminosity. 

The case of the collapsing star is more challenging and has been
only studied previously either assuming an electrovacuum magnetosphere
and/or by matching the exterior to the interior solution. Our results
are qualitatively similar to those found in~\citep{2011arXiv1112.2622L},
although the total radiated energy in our simulations is one order of magnitude
larger due to an increase in both the peak energy in the magnetosphere
and the fraction of radiated energy. The possible detectability
of this burst has been already discussed in detail in~\citep{2011arXiv1112.2622L}
and therefore will not be repeated here.

In conclusion, the resistive MHD framework allows to
consider a broad range of new phenomena;study reconnections and dissipation
with more realistic Ohm's law - like the resistive solutions of
pulsar magnetospheres~\citep{2012ApJ...746...60L}-, model the magnetic growth due
to different instabilities by using the mean-field dynamo~\citep{2012arXiv1205.2951B},
and compute the magnetosphere interaction of binary systems --like neutron-neutron
stars and neutron-black hole--, which may be crucial to study the possible
electromagnetic counterparts to the gravitational waves emitted by these systems,
among others possibilities. Work on these directions is in progress and it will be
reported in the near future.

%%%%%%%%%%%%%%%%%%%%%%%%%%%%%%%%%%%%%%%%%%%%%%%
\appendix

%%%%%%%%%%%%%%%%%%%%%%%%%%%%%%%%%%%%%%%%%%%%%%%

\section{IMEX}\label{appendixA}
%%%%%%%%%%%%%%%%%%%%%%%%%%%%%%%%%%%%%%%%%%%%%%%

IMEX Runge-Kutta schemes can be represented by a double tableau in
the usual Butcher notation~\citep{But:1987,But:2003}

\begin{equation}
\begin{minipage}{1.2in}
\begin{tabular} {c c c}
${\tilde c}$  & \vline & ${\tilde A}$  \\
\hline 
              & \vline & ${\tilde \omega}^T$  
\end{tabular}
\end{minipage} 
\hskip 1.0cm
\begin{minipage}{1.2in}
\begin{tabular} {c c c}
${c}$  &  \vline & ${A}$  \\
\hline 
       &  \vline & ${\omega}^T$  
\label{butcher_tableau}
\end{tabular}
\end{minipage}
\end{equation}

where the coefficients $\tilde{c}$ and $c$ used for the treatment
of non-autonomous systems are given by the following relation
\begin{equation}\label{definition_cs}
   {\tilde c}_{i} = \sum_{j=1}^{i-1}~ {\tilde{a}}_{ij} ~~~,~~~
   {c}_{i} = \sum_{j=1}^{i}~ {a}_{ij} ~~~.
\end{equation}

Solutions of conservation equations have some norm that decreases in time.
It would be desirable, in order to avoid spurious numerical oscillations
arising near discontinuities of the solution, to maintain such property
at a discrete level by the numerical method. The most commonly used norms
are the TV-norm and the infinity norm. A scheme is called Strong Stability
Preserving (SSP) if maintains a given norm during the
evolution~\citep{SpiRuu:2002}.

In all these schemes the implicit tableau corresponds to an L-stable scheme 
(that is, $\omega^T A^{-1} e =1$, being $e$ a vector whose components are
all equal to $1$), whereas the explicit tableau is SSP$k$, where $k$ denotes
the order of the SSP scheme. We shall use the notation SSP$k(s,\sigma,p)$,
where the triplet $(s,\sigma,p)$ characterizes the number of $s$ stages
of the implicit scheme, the number $\sigma$ of stages of the explicit scheme
and the order $p$ of the IMEX scheme.

There are different IMEX RK schemes available in the literature. We have
considered only third order IMEX schemes, some of them found in the
literature~\citep{ParRus:2005} and others developed by us. All of them are
based on a third order SSP explicit scheme that can be implemented
efficiently by using only two levels of fields and one of rhs.
It is worth mentioning that these methods are still under development
and have few drawbacks. Probably the most serious one is an accuracy
degradation for some range of the relaxation time $\epsilon$.

\begin{table*}
\caption{Tableau for the explicit (left) implicit (right) IMEX-SSP3(4,3,3)
L-stable scheme}
\begin{minipage}{1.8in}
\begin{tabular} {c c c c c c}
 0   & \vline & 0  &  0  &  0  & 0  \\
 0   & \vline & 0  &  0  &  0  & 0  \\
 1   & \vline & 0  &  1  &  0  & 0 \\
 1/2 & \vline & 0  & 1/4 & 1/4 & 0 \\
\hline 
   & \vline &  0 & 1/6 & 1/6 & 2/3 \\
\end{tabular}
\end{minipage}
\begin{minipage}{1.8in}
\begin{tabular} {c c c c c c}
  $\alpha$   & \vline & $\alpha$  &  0  &  0  & 0  \\
 0   & \vline & -$\alpha$  &  $\alpha$  &  0  & 0  \\
 1   & \vline & 0  &  $1-\alpha$  &  $\alpha$  & 0 \\
 1/2 & \vline & $\beta$  & $\eta$ & $1/2-\beta-\eta-\alpha$ & $\alpha$ \\
\hline 
   & \vline &  0 & 1/6 & 1/6 & 2/3 \\
\end{tabular}
\end{minipage}
\begin{equation}
 \alpha = 0.24169426078821~,~\beta = 0.06042356519705~,~ 
 \eta = 0.12915286960590 \nonumber
\end{equation}
\label{SSP3-433}
\end{table*}

\begin{table*}
\caption{Tableau for the explicit (left) implicit (right) IMEX-SSP3(5,3,3)
L-stable scheme}
\begin{minipage}{1.8in}
\begin{tabular} {c c c c c c c}
 0   & \vline & 0  &  0  &  0  & 0 & 0 \\
 0   & \vline & 0  &  0  &  0  & 0 & 0 \\
 1   & \vline & 0  &  1  &  0  & 0 & 0 \\
 1/2 & \vline & 0  & 1/4 & 1/4 & 0 & 0 \\
  1  & \vline & 0  & 1/6 & 1/6 & 2/3 & 0 \\
\hline 
   & \vline &  0 & 1/6 & 1/6 & 2/3 & 0 \\
\end{tabular}
\end{minipage}
\hskip 1.0cm
\begin{minipage}{1.8in}
\begin{tabular} {c c c c c c c}
  $\alpha$   & \vline & $\alpha$  &  0  &  0  & 0  & 0\\
 0   & \vline & -$\alpha$  &  $\alpha$  &  0  & 0  & 0 \\
 1   & \vline & 0  &  $1-\alpha$  &  $\alpha$  & 0 & 0\\
 1/2 & \vline & $a_{41}$  & $a_{42}$ & $a_{43}$ & $\alpha$ & 0\\
  1 & \vline &  0 & 1/6 & 0 & 2/3 & 1/6 \\
\hline 
   & \vline &  0 & 1/6 & 0 & 2/3 & 1/6\\
\end{tabular}
\end{minipage}
\begin{equation}
 a_{41} = \frac{1}{8 \alpha} (2 \alpha^2 + 2 \alpha -1) ~~,~~
 a_{42} = \frac{1}{8 \alpha} (-4 \alpha^2 + 1) ~~,~~ 
 a_{43} =  \frac{1}{4} (-3 \alpha + 1) ~~,~~
 \alpha = 1/3~~. \nonumber 
\end{equation}
\label{SSP3-533}
\end{table*}

%%%%%%%%%%%%%%%%%%%%%%%%%%%%%%%%%%%%%%%%%%%%%%%
\section{Ideal MHD limit}\label{appendixB}
%%%%%%%%%%%%%%%%%%%%%%%%%%%%%%%%%%%%%%%%%%%%%%%

The ideal MHD limit can be obtained by requiring the current
to be finite even in the limit of infinite isotropic conductivity,
leading to the condition $E^i = - \epsilon^{ijk} v_j B_k$.
The Ohm's law current becomes undetermined (i.e., an infinite
conductivity multiplying a vanishing electric field in the co-moving frame),
but it can still be computed from the redundant Maxwell 
equation for the electric field evolution (\ref{maxwellext_3+1_eq1a}).
The evolution of the magnetic field can be simplified by substituting
the ideal MHD condition in (\ref{maxwellext_3+1_eq1c}),
\begin{eqnarray}\label{idealmhd_B}
  \partial_t (\sqrt{\gamma} B^{i}) &+& 
  \partial_k [\sqrt{\gamma} \{ (\alpha v^k - \beta^k) B^i
         - \alpha v^i B^k + \alpha \gamma^{ki} \phi \}] \nonumber \\
   &=&   \sqrt{\gamma} [-B^k \partial_k \beta^i 
         + \phi \gamma^{ik} (\partial_k \alpha + \Gamma^j_{jk})]
\end{eqnarray}

The transformation from conserved to primitive is simplified by 
eliminating the electric field as an independent variable
and may allow us to recover the primitive quantities in a more robust way.
Substituting the ideal MHD condition in the definition of the conserved
variables
\begin{eqnarray}
   \tau &=& h W^2 + B^2 - p  - D - \frac{1}{2} [(B^k v_k)^2 + \frac{B^2}{W^2}] ~~~,~~~ 
\label{Tmunu_decomposition2a} \\
   S_{i} &=& [h W^2 + B^2] v_{i} - (B^k v_k) B_i ~~~.
\label{Tmunu_decomposition2b}
\end{eqnarray}
it is easy to check that
\begin{equation}\label{con2prim_mhd_vB}
     v_i B^i = \frac{S_i B^i}{h W^2} ~~.
\end{equation}
Using this relation, the scalar product $S^i S_i$ can be solved for
the Lorentz factor, obtaining
\begin{equation}\label{con2prim_mhd1}
   c \equiv \frac{1}{W^2} = 1 - \frac{x^2 S^2 + (2 x + B^2)(S_i B^i)^2}{x^2 (x + B^2)^2}
\end{equation}

Assuming an ideal gas EoS, and after some manipulations in the
definition of $\tau$ (\ref{Tmunu_decomposition2a}), 
the resulting final equation to solve is
\begin{eqnarray}\label{trascendental_idealMHD}
   f(x) &=& [ 1 - \frac{(\Gamma - 1) c}{\Gamma}] x
   + [\frac{(\Gamma-1) \sqrt{c}}{\Gamma} - 1] D 
\nonumber \\
   &+& [1-\frac{c}{2}] B^2 - \frac{1}{2 x^2} (S_i B^i)^2 - \tau ~~.
\end{eqnarray}

%%%%%%%%%%%%%%%%%%%%%%%%%%%%%%%%%%%%%%%%%%%%%%%%%%%%%%%%%%%%%%%%%%%%
%
%   B I B L I O G R A P H Y
%
%%%%%%%%%%%%%%%%%%%%%%%%%%%%%%%%%%%%%%%%%%%%%%%%%%%%%%%%%%%%%%%%%%%%
%%%%%%%%%%%%%%%%%%%%%%%%%%%%%%%%%%%%%%%%%%%%%%%%%%%%%%%%%%%%%%
\section*{Acknowledgments}
%%%%%%%%%%%%%%%%%%%%%%%%%%%%%%%%%%%%%%%%%%%%%%%%%%%%%%%%%%%%%%

The author acknowledges his long time collaborators 
E~.Hirschmann, S.~Liebling and C~.Thompson for useful comments,
and particularly to D.~Alic for discussions on the matching
of the currents, D.~Neilsen for his help on implementing the
IMEX in HAD, and L.~Lehner for carefully reading and discussing
this manuscript. This work was supported by the Jeffrey
L.~Bishop Fellowship. Computations were performed in Scinet.

\bibliographystyle{mn2e}
%\bibliography{rmhd}

\begin{thebibliography}{50}
\expandafter\ifx\csname natexlab\endcsname\relax\def\natexlab#1{#1}\fi

\bibitem[{{Alic} {et~al.}(2012){Alic}, {Moesta}, {Rezzolla}, {Zanotti}, \&
  {Jaramillo}}]{2012ApJ...754...36A}
{Alic} D., {Moesta} P., {Rezzolla} L., {Zanotti} O., {Jaramillo} J.~L., 2012,
  \apj, 754, 36

\bibitem[{Anderson {et~al.}(2006)Anderson, Hirschmann, Liebling, \&
  Neilsen}]{Anderson:2006ay}
Anderson M., Hirschmann E., Liebling S.~L., Neilsen D., 2006, Class. Quant.
  Grav., 23, 6503

\bibitem[{Anderson {et~al.}(2008)}]{Anderson:2007kz}
Anderson M., {et~al.}, 2008, Phys. Rev., D77, 024006

\bibitem[{{Andersson}(2012)}]{2012arXiv1204.2695A}
{Andersson} N., 2012, ArXiv e-prints

\bibitem[{{Balbus} \& {Hawley}(1991)}]{1991ApJ...376..214B}
{Balbus} S.~A., {Hawley} J.~F., 1991, \apj, 376, 214

\bibitem[{{Balbus} \& {Hawley}(1998)}]{1998RvMP...70....1B}
---, 1998, Reviews of Modern Physics, 70, 1

\bibitem[{Baumgarte \& Shapiro(2003)}]{Baumgarte:2002b}
Baumgarte T.~W., Shapiro S.~L., 2003, Astrophys. J., 585, 930

\bibitem[{{Bekenstein} \& {Oron}(1978)}]{1978PhRvD..18.1809B}
{Bekenstein} J.~D., {Oron} E., 1978, \prd, 18, 1809

\bibitem[{{Blandford} \& {Znajek}(1977)}]{Blandford1977}
{Blandford} R.~D., {Znajek} R.~L., 1977, Mon. Not. R. Astron. Soc., 179, 433

\bibitem[{{Bucciantini} \& {Del Zanna}(2012)}]{2012arXiv1205.2951B}
{Bucciantini} N., {Del Zanna} L., 2012, ArXiv e-prints

\bibitem[{{Bucciantini} {et~al.}(2006){Bucciantini}, {Thompson}, {Arons},
  {Quataert}, \& {Del Zanna}}]{2006MNRAS.368.1717B}
{Bucciantini} N., {Thompson} T.~A., {Arons} J., {Quataert} E., {Del Zanna} L.,
  2006, \mnras, 368, 1717

\bibitem[{Butcher(1987)}]{But:1987}
Butcher J., 1987

\bibitem[{Butcher(2003)}]{But:2003}
---, 2003

\bibitem[{{Campanelli} {et~al.}(2006){Campanelli}, {Lousto}, {Marronetti}, \&
  {Zlochower}}]{2006PhRvL..96k1101C}
{Campanelli} M., {Lousto} C.~O., {Marronetti} P., {Zlochower} Y., 2006,
  Physical Review Letters, 96, 111101

\bibitem[{{Contopoulos} \& {Spitkovsky}(2006)}]{2006ApJ...643.1139C}
{Contopoulos} I., {Spitkovsky} A., 2006, \apj, 643, 1139

\bibitem[{{Dedner} {et~al.}(2002){Dedner}, {Kemm}, {Kr{\"o}ner}, {Munz},
  {Schnitzer}, \& {Wesenberg}}]{2002JCoPh.175..645D}
{Dedner} A., {Kemm} F., {Kr{\"o}ner} D., {Munz} C.-D., {Schnitzer} T.,
  {Wesenberg} M., 2002, Journal of Computational Physics, 175, 645

\bibitem[{{Dionysopoulou} {et~al.}(2012){Dionysopoulou}, {Alic}, {Palenzuela},
  {Rezzolla}, \& {Giacomazzo}}]{2012arXiv1208.3487D}
{Dionysopoulou} K., {Alic} D., {Palenzuela} C., {Rezzolla} L., {Giacomazzo} B.,
  2012, ArXiv e-prints

\bibitem[{{Dumbser} \& {Zanotti}(2009)}]{2009JCoPh.228.6991D}
{Dumbser} M., {Zanotti} O., 2009, Journal of Computational Physics, 228, 6991

\bibitem[{Goldreich \& Julian(1969)}]{Goldreich:1969sb}
Goldreich P., Julian W.~H., 1969, Astrophys.J., 157, 869

\bibitem[{{Gruzinov}(2007)}]{2007ApJ...667L..69G}
{Gruzinov} A., 2007, \apjl, 667, L69

\bibitem[{{Hawley} \& {Balbus}(1991)}]{1991ApJ...376..223H}
{Hawley} J.~F., {Balbus} S.~A., 1991, \apj, 376, 223

\bibitem[{{Hawley} {et~al.}(1995){Hawley}, {Gammie}, \&
  {Balbus}}]{1995ApJ...440..742H}
{Hawley} J.~F., {Gammie} C.~F., {Balbus} S.~A., 1995, \apj, 440, 742

\bibitem[{{Kalapotharakos} \& {Contopoulos}(2009)}]{2009A&A...496..495K}
{Kalapotharakos} C., {Contopoulos} I., 2009, \aap, 496, 495

\bibitem[{{Komissarov}(2004)}]{2004MNRAS.350..427K}
{Komissarov} S.~S., 2004, \mnras, 350, 427

\bibitem[{{Komissarov}(2007)}]{Komissarov2007}
---, 2007, Mon. Not. R. Astron. Soc., 382, 995

\bibitem[{Lehner {et~al.}(2006)Lehner, Liebling, \& Reula}]{Lehner:2005vc}
Lehner L., Liebling S.~L., Reula O., 2006, Class. Quant. Grav., 23, S421

\bibitem[{{Lehner} {et~al.}(2011){Lehner}, {Palenzuela}, {Liebling},
  {Thompson}, \& {Hanna}}]{2011arXiv1112.2622L}
{Lehner} L., {Palenzuela} C., {Liebling} S.~L., {Thompson} C., {Hanna} C.,
  2011, ArXiv e-prints

\bibitem[{{Li} {et~al.}(2012){Li}, {Spitkovsky}, \&
  {Tchekhovskoy}}]{2012ApJ...746...60L}
{Li} J., {Spitkovsky} A., {Tchekhovskoy} A., 2012, \apj, 746, 60

\bibitem[{Liebling(2002)}]{Liebling}
Liebling S.~L., 2002, Phys. Rev. D, 66, 041703

\bibitem[{{Lyutikov}(2011)}]{2011PhRvD..83l4035L}
{Lyutikov} M., 2011, \prd, 83, 124035

\bibitem[{McKinney(2006)}]{McKinney:2006sd}
McKinney J.~C., 2006, Mon. Not. Roy. Astron. Soc. Lett., 368, L30

\bibitem[{{Moesta} {et~al.}(2012){Moesta}, {Alic}, {Rezzolla}, {Zanotti}, \&
  {Palenzuela}}]{2012ApJ...749L..32M}
{Moesta} P., {Alic} D., {Rezzolla} L., {Zanotti} O., {Palenzuela} C., 2012,
  \apjl, 749, L32

\bibitem[{{Neilsen} {et~al.}(2011){Neilsen}, {Lehner}, {Palenzuela},
  {Hirschmann}, {Liebling}, {Motl}, \& {Garrett}}]{2011PNAS..10812641N}
{Neilsen} D., {Lehner} L., {Palenzuela} C., {Hirschmann} E.~W., {Liebling}
  S.~L., {Motl} P.~M., {Garrett} T., 2011, Proceedings of the National Academy
  of Science, 108, 12641

\bibitem[{Newman \& Penrose(1962)}]{Newman:1961qr}
Newman E., Penrose R., 1962, J.Math.Phys., 3, 566

\bibitem[{{Obergaulinger} {et~al.}(2010){Obergaulinger}, {Aloy}, \&
  {M{\"u}ller}}]{2010A&A...515A..30O}
{Obergaulinger} M., {Aloy} M.~A., {M{\"u}ller} E., 2010, \aap, 515, A30

\bibitem[{{Palenzuela} {et~al.}(2011){Palenzuela}, {Bona}, {Lehner}, \&
  {Reula}}]{2011CQGra..28m4007P}
{Palenzuela} C., {Bona} C., {Lehner} L., {Reula} O., 2011, Classical and
  Quantum Gravity, 28, 134007

\bibitem[{{Palenzuela} {et~al.}(2010{\natexlab{a}}){Palenzuela}, {Garrett},
  {Lehner}, \& {Liebling}}]{2010PhRvD..82d4045P}
{Palenzuela} C., {Garrett} T., {Lehner} L., {Liebling} S.~L.,
  2010{\natexlab{a}}, \prd, 82, 044045

\bibitem[{{Palenzuela} {et~al.}(2010{\natexlab{b}}){Palenzuela}, {Lehner}, \&
  {Liebling}}]{2010Sci...329..927P}
{Palenzuela} C., {Lehner} L., {Liebling} S.~L., 2010{\natexlab{b}}, Science,
  329, 927

\bibitem[{{Palenzuela} {et~al.}(2009){Palenzuela}, {Lehner}, {Reula}, \&
  {Rezzolla}}]{2009MNRAS.394.1727P}
{Palenzuela} C., {Lehner} L., {Reula} O., {Rezzolla} L., 2009, \mnras, 394,
  1727

\bibitem[{{Palenzuela} {et~al.}(2010{\natexlab{c}}){Palenzuela}, {Lehner}, \&
  {Yoshida}}]{2010PhRvD..81h4007P}
{Palenzuela} C., {Lehner} L., {Yoshida} S., 2010{\natexlab{c}}, \prd, 81,
  084007

\bibitem[{Pareschi \& Russo(2005)}]{ParRus:2005}
Pareschi L., Russo G., 2005, J. Sci. Comput., 25, 112

\bibitem[{Pretorius(2002)}]{Pretoriusphd}
Pretorius F., 2002, PhD thesis, The University of British Columbia

\bibitem[{{Price} \& {Rosswog}(2006)}]{2006Sci...312..719P}
{Price} D.~J., {Rosswog} S., 2006, Science, 312, 719

\bibitem[{Spiteri \& Ruuth(2002)}]{SpiRuu:2002}
Spiteri R., Ruuth S., 2002, SIAM J. Numer. Anal., 40(2), 469

\bibitem[{{Spitkovsky}(2006)}]{2006ApJ...648L..51S}
{Spitkovsky} A., 2006, \apjl, 648, L51

\bibitem[{{Takamoto} \& {Inoue}(2011)}]{2011ApJ...735..113T}
{Takamoto} M., {Inoue} T., 2011, \apj, 735, 113

\bibitem[{{Tchekhovskoy} \& {Spitkovsky}(2012)}]{2012arXiv1211.2803T}
{Tchekhovskoy} A., {Spitkovsky} A., 2012, ArXiv e-prints

\bibitem[{{Uzdensky}(2011)}]{2011SSRv..160...45U}
{Uzdensky} D.~A., 2011, \ssr, 160, 45

\bibitem[{{Zanotti} \& {Dumbser}(2011)}]{2011MNRAS.418.1004Z}
{Zanotti} O., {Dumbser} M., 2011, \mnras, 418, 1004

\bibitem[{{Zenitani} {et~al.}(2010){Zenitani}, {Hesse}, \&
  {Klimas}}]{2010ApJ...716L.214Z}
{Zenitani} S., {Hesse} M., {Klimas} A., 2010, \apjl, 716, L214

\end{thebibliography}

%%%%%%%%%%%%%%%%%%%%%%%%%%%%%%%%%%%%%%%%%%%%%%%%%%%%%%%%%%%%%%%%%%%%
%
%   E N D   D O C U M E N T
%
%%%%%%%%%%%%%%%%%%%%%%%%%%%%%%%%%%%%%%%%%%%%%%%%%%%%%%%%%%%%%%%%%%%%
\label{lastpage}

\end{document}